\documentclass[%
 aip,
 jcp,
 amsmath,amssymb,
 reprint,%
]{revtex4-1}
\usepackage{dcolumn,booktabs}
\newcolumntype{d}[1]{D{.}{.}{#1}}

\usepackage{graphicx}
\usepackage{dcolumn}
\usepackage{bm}

\usepackage{epstopdf}

\usepackage[utf8]{inputenc}
\usepackage[T1]{fontenc}
\usepackage{mathptmx}
\usepackage{physics}
\usepackage{subfig}
\usepackage{color}
\usepackage{xcolor} 

\newcommand{\luuk}[1]{{\textcolor{orange}{Luuk: #1 }} }

\begin{document}

\preprint{AIP/123-QED}

\title[DMRG-tailored coupled cluster method in the 4c-relativistic domain]{DMRG-tailored coupled cluster method in the 4c-relativistic domain: General implementation and application to the NUHFI and NUF$_3$ molecules}

\author{Jakub Vi{\v s}{\v n}\'ak}
\affiliation{J. Heyrovsk\'{y} Institute of Physical Chemistry, Academy of Sciences of the Czech \mbox{Republic, v.v.i.}, Dolej\v{s}kova 3, 18223 Prague 8, Czech Republic}
\affiliation{Faculty of Mathematics and Physics, Charles University, Prague, Czech Republic}
\affiliation{Middle East Technical University, Üniversiteler Mahallesi, Dumlupınar Bulvarı No:1, 06800 Çankaya Ankara, Türkiye}

\author{Jan Brandejs}
\affiliation{J. Heyrovsk\'{y} Institute of Physical Chemistry, Academy of Sciences of the Czech \mbox{Republic, v.v.i.}, Dolej\v{s}kova 3, 18223 Prague 8, Czech Republic}
\affiliation{Faculty of Mathematics and Physics, Charles University, Prague, Czech Republic}
\affiliation{Faculty of Science, Humanities and Education, Technical University of Liberec, Czech Republic}
\author{Mih{\'a}ly M{\'a}t{\'e}}
\affiliation{Strongly Correlated Systems ``Lend{\"u}let'' Research Group,
Wigner Research Centre for Physics,
H-1121 Budapest, Konkoly-Thege Mikl{\'o}s {\'u}t 29-33, Hungary}
\affiliation{Department of Mathematics, Technical University of Munich, Germany}

\author{Lucas Visscher}
\affiliation{Department of Chemistry and Pharmaceutical Sciences, De Boelelaan 1108, Vrije Universiteit Amsterdam, NL-1081 HZ Amsterdam, Netherlands}

\author{\"Ors Legeza}
\email{legeza.ors@wigner.mta.hu}
\affiliation{Strongly Correlated Systems ``Lend{\"u}let'' Research Group,
Wigner Research Centre for Physics,
H-1121 Budapest, Konkoly-Thege Mikl{\'o}s {\'u}t 29-33, Hungary}
\affiliation{Institute for Advanced Study, Technical University of Munich, Lichtenbergstrasse 2a, 85748 Garching, Germany}
\author{Ji{\v r}{\'i} Pittner}
\email{jiri.pittner@jh-inst.cas.cz}
\affiliation{J. Heyrovsk\'{y} Institute of Physical Chemistry, Academy of Sciences of the Czech \mbox{Republic, v.v.i.}, Dolej\v{s}kova 3, 18223 Prague 8, Czech Republic}

\date{\today}

\begin{abstract}
Heavy atom compounds represent a challenge for computational chemistry, due to the need for simultaneous treatment of relativistic and correlation effects. Often such systems exhibit also strong correlation which hampers the application
of perturbation theory or single-reference coupled cluster (CC) methods. As a viable alternative, we have proposed to externally correct the CC method using the density matrix renormalization group (DMRG) wave functions, yielding the DMRG-tailored CC method.
In a previous paper [J. Chem. Phys. {\bf 152}, 174107 (2020)] we have reported a first implementation of this method in the relativistic context, which was restricted to molecules with real double group symmetry. In this work we present a fully general implementation of the method, covering complex and quaternion double groups as well. The 4c-TCC method thus becomes applicable to polyatomic molecules including heavy atoms. For assessment of the method, we performed calculations of the chiral uranium compound NUHFI,
which was previously studied in the context of the enhancement of parity violation effects. In particular, we performed
calculations of a cut of the potential energy surface of this molecule along the dissociation of the N-U bond, where the
system exhibits a strong multireference character.
Since there are no experimental data for NUHFI, we have performed also an analogous study of the (more symmetric) NUF$_3$ molecule, where the vibrational frequency of the N-U bond can be compared with spectroscopic data.
\end{abstract}

\maketitle
\section{INTRODUCTION}

The density matrix renormalization group method (DMRG) \cite{White1992} was introduced to quantum-chemical community \cite{White1999, Chan2002, Legeza2003a} about two decades ago and since then, it has seen many applications to multireference systems and the DMRG-SCF method is becoming a standard computational approach replacing traditional CASSCF. 
The biggest advantage of the DMRG method is its capability to treat large active spaces, current implementations can go beyond 50 active space spinors 
\cite{Schollwoeck2011, OlivaresAmaya2015}.
However, a major drawback of DMRG is its inability to capture dynamical correlation, since it cannot include all virtual spinors.  The dynamic correlation plays an important role in virtually all molecules and its accurate description is important when trying to achieve chemical accuracy.
Although the DMRG method is already well established, the methods for treating the dynamical correlations on top of DMRG are still in pioneering stage. Past efforts were either based on second order perturbation theory \cite{Kurashige2011}, internally contracted MRCI (multireference configuration interaction) \cite{Saitow2013}, random phase approximation \cite{Wouters2013}, canonical transformation method \cite{Yanai2006}, the perturbation theory with matrix product states \cite{Ren2016}, the adiabatic connection method \cite{Veis-Pernal2021}, and the restricted active space density matrix renormalization group method \cite{Barcza-2022b,Friesecke-2023}.

In our group we employed an approach based on the coupled cluster method externally corrected by DMRG \cite{Veis2016}. As the name suggests, this is a combination of DMRG and the coupled cluster (CC) method. The CC method is well known for its efficiency in describing dynamical correlation. In the externally corrected approach, first a DMRG calculation is performed within the strongly correlated active space fixed which accounts for the static correlation. Second step is the CC analysis of the matrix product state (MPS) wave function, obtained from DMRG
which yields CC aplitudes with indices restricted to the active space spinorbitals. Then a CC calculation is performed on the whole orbital space, keeping the active space amplitudes fixed, which captures the dynamical correlation.
Already the simplest version thereof, the tailored CCSD (CC with single and double excitations) approach \cite{Kinoshita2005,Hino2006}, yields very promising results \cite{Veis2016,Veis2016err,Faulstich2019a,Faulstich2019b,Leszczyk-2022}. In the non-relativistic context,
it has been implemented in the Orca program\cite{Neese2012}, employing the domain-based local pair natural orbital (DLPNO) 
techniques to achieve high computational efficiency \cite{lpno-tcc,dlpno-tcc,dlpno-tcc-app}.


In a previous paper \cite{reltcc2020} we have presented the DMRG-tailored coupled cluster method in the 4c-relativistic domain
in an implementation restricted to real double groups, i.e. basically to diatomic or other linear molecules. Here we report a general implementation
applicable to polyatomic molecules with complex or quaternion double group symmetry.

Computational studies of polyatomic molecules with low symmetry at the four-component relativistic and post-HF correlated level are relatively scarce in the literature. To illustrate the applicability of our approach, we have 
selected the NUHFI molecule, which is a chiral compound and was previously studied at the relativistic X2C-DFT level in the context of the enhancement of parity violation effects \cite{Wormit2014}. 
For the search for parity violation effects in high-resolution spectroscopy experiments it is convenient if the molecule
has a vibrational frequency in the range of the CO$_2$ laser (900-1100 cm$^{-1}$), which corresponds to the stretch
of the U$\equiv$N triple bond in this molecule. On the other hand, stretching a triple bond beyond the immediate
vicinity of equlibrium quicky introduces appreciable non-dynamic correlation, and the N$_2$ molecule has been
often employed as a benchmark to test performance of multireference correlated methods \cite{Jiang2011,Mintz2009,Chan-2004b,Faulstich2019b,Mate-2022} and to demonstrate
the failure of single-reference ones to properly describe dissociation of this bond \cite{LiPaldus1998}. This has motivated
us to choose the NUHFI molecule as a relativistic variant of such a benchmark and to study the cut of the
potential energy surface of this molecule along the stretch of the N$\equiv$U bond. Since there are no experimental data available for this molecule, we have also performed a study of the NUF$_3$ molecule, which has an analogous N-U bond and where
vibrational frequency and anharmonicity has been measured.
However, due to its higher symmetry, this molecule does not fully exploit our most general implementation of the 4c-TCCSD method.

\section{THEORY AND IMPLEMENTATION\label{theorysection}}

The state-of-the-art post-HF relativistic calculations employ the no-virtual-pair approximation
and the empty Dirac picture, i.e. the negative-energy solutions of the Dirac-Coulomb Hamiltonian are projected out
and only the electronic solutions enter the correlated calculation. No excitations to the positronic manifold thus occurs, corresponding to the neglect of virtual electron-positron pair creation. This is a well substantiated approximation at the chemical energy scale and it yields a second-quantized Hamiltonian formally analogous to the non-relativistic case
\begin{eqnarray}
\label{ham2q}
H = \sum_{PQ} h_P^Q \, a_P^\dagger a_Q + \frac{1}{4}  \sum_{PQRS} \langle PQ || RS\rangle \, a^\dagger_P a^\dagger_Q a_S a_R,
\end{eqnarray}
where the indices $P,Q,R,S$ run over the positive-energy 4-component spinors spanning the one-electron basis.
In the Kramers-restricted formalism barred spinors ($\phi_{\bar{p}}$) and unbarred spinors ($\phi_{p}$) form  Kramers pairs related to each other by action of the time-reversal operator $K$
\begin{eqnarray}
\label{kramers}
K \phi_p &=& \phi_{\bar{p}}, \nonumber \\
K \phi_{\bar{p}} &=& -\phi_p.
\end{eqnarray}
The Kramers symmetry, which leads to a twofold degeneracy of spinor energies in the absence of magnetic fields\cite{KRAMERS1930}, can be used to play the role that spin symmetry has in non-relativistic theory. In such approaches the $M_S$  good quantum number
is replaced by the $M_K$ quasi quantum number\cite{fleig2001}, which is $1/2$ for unbarred spinors (A) and $-1/2$ for spinors with barred indices (B).
The capital indices in (\ref{ham2q}) run over both A and B components of Kramers pairs.
In contrast to the non-relativistic case, the Hamiltonian (\ref{ham2q}) is in general not block-diagonal in $M_K$, but this number can then serve to partition the operators and Hamiltonian blocks in much the same way as is done with the $M_S$ quantum number.
Since each creation or annihilation operator in (\ref{ham2q}) changes $M_K$ by $\pm 1/2$, the Hamiltonian can only directly couple states with $|\Delta M_K| \leq 2$. States with a still higher difference in $|M_K|$ are only coupled indirectly and can possibly be neglected if the Kramers' pairing is chosen such that it minimizes the couplings between states with different $M_K$ values. 
Also with the aid of Kramers' symmetry, the index permutation symmetry of the 2e-integrals in (\ref{ham2q}) is still lower than in the non-relativistic case as this is based on the use of real-valued spin-orbitals, whereas relativistic spinors are in general complex-valued.

The Dirac program \cite{DIRAC18} employs a quaternion symmetry approach which combines the Kramers and binary double group
symmetry ($D_{2h}^*$ and subgroups) \cite{saue-jensen-1999}.
The binary double groups can be divided into three classes based on the Frobenius-Schur indicator:
``real groups'' ($D_{2h}^*$, $D_{2}^*$, and $C_{2v}^*$);
``complex groups'' ($C_{2h}^*$, $C_{2}^*$, and $C_{s}^*$); 
and ``quaternion groups'' ($C_i^*$ and $C_1^*$) \cite{dyall-faegri}.
Generalization of non-relativistic post-HF methods is simplest for the ``real groups'', 
where the integrals in (\ref{ham2q}) are real-valued and the ones with odd number of barred (B) indices vanish,
i.e. one has only to include additional ``spin cases'' of integrals $(\mathrm{AB}|\mathrm{AB})$ and $(\mathrm{AB}|\mathrm{BA})$ (in Mulliken notation) 
as well as take care of the fact that in general integrals $(\mathrm{BB}|\mathrm{AA}) \neq (\mathrm{AA}|\mathrm{AA})$.
For the complex groups, the integrals are complex-valued, but still only integrals with an even number of barred indices are non-zero.
Finally, in the lowest symmetry case of  ``quaternion groups'' all the integrals have to be included
and are complex-valued \cite{Visscher1996,Visscher2002,Thyssen_phd, dyall-faegri}.

The tailored coupled cluster (CC) method belongs to the broader category of externally corrected CC methods,
which take information on static correlation from some non-CC external source, and  include it into
the subsequent CC treatment \cite{paldus-externalcorr}.
Tailored CC method (TCC),
proposed by Bartlett et al.\cite{Kinoshita2005,cyclobut-tailored-2011,melnichuk-2012,melnichuk-2014}, 
is the conceptually simplest such method. It employes the split-amplitude ansatz for
the wave function introduced by Piecuch et al. \cite{Piecuch1993,semi3}
\begin{align}
\ket{\Psi} = e^{T_{\mathrm{ext}}} e^{T_{\mathrm{cas}}} \ket{ \Phi}
        \label{eqn:ansatz}
\end{align}
where $\ket{\Phi}$ is the Dirac-Hartree-Fock Slater determinant and $T_{\mathrm{cas}}$ containing amplitudes with all active indices is ``frozen'' at values obtained from CASCI or in our case from DMRG.
The external cluster operator $T_{\mathrm{ext}}$ is composed of amplitudes with at least one index outside the DMRG active space.

The simplest version of the method truncates  both  $T_{\mathrm{cas}}$  and  $T_{\mathrm{ext}}$ to single and double excitations.
The excitation operators $T_{\mathrm{ext}}$ and $T_{\mathrm{cas}}$ commute,
which allows to
use the standard CCSD solver, modified to keep the amplitudes from $T_{\mathrm{cas}}$ fixed
at values obtained from the DMRG MPS wave function \cite{Legeza2003}.
Since the Hamiltonian contains only one- and two-body terms, the TCCSD energy with the $T_{\mathrm{ext}}=0$ and $T_{\mathrm{cas}}$ from DMRG reproduces the DMRG energy.
In the limit of the DMRG active space including all MOs, TCC thus recovers the FCI energy.
In general, a quadratic error bound valid for TNS-TCC methods has been derived \cite{Faulstich2019a}.

In Refs.\cite{Veis2016,Veis2016err} we have presented an implementation of the TCCSD method for the non-relativistic case,
which was later followed by more efficient LPNO and DLPNO based implementations \cite{lpno-tcc,dlpno-tcc}, and
the implementation of TCCSD for the simpler real-group symmetry case followed \cite{reltcc2020}.
In contrast to the non-relativistic DMRG, where the ``sites'' correspond to spatial orbitals with a 4-dimensional
site basis, in the relativistic case\cite{Battaglia2018} a site represents a single Kramers spinor, being occupied or unoccupied.
The number of sites for the DMRG sweeps thus doubles with respect to the non-relativistic case.  In general, it is also possible to form a 4-dimensional ``super site'' by taking the tensor product basis
of the barred and unbarred Kramers spinors which might look more beneficial by reducing the number of sites by two. However, in this implementation the quantum number decomposed site operators ~\cite{Schollwoeck2011,Szalay2015} do not reduce to scalar multiplicative factors via the diagonalization procedure and consequently the
corresponding general tensor algebra turns out to be computationally more demanding.
Extraction of the CI coefficients from the MPS wave function also needs an optimal implementation for system sizes
in the range of 100 sites or more. As there is a large redundancy in occupation number ``strings'', i.e. arrays with binary elements corresponding to the various determinants, we form several precontracted components for the left 
and right DMRG subsystem MPS components, thus the final contraction of the network can be accelerated tremendously. In addition, these can be performed very efficiently via massive parallelization. Our code can also handle arbitrary Abelian and non-Abelian symmetries which again could reduce computational time and memory requirement significantly.


From the technical point of view, in our previous works, outlined above, the Budapest DMRG code has been interfaced with the Dirac code allowing us to perform the TCC workflow for heavy diatomic molecules when relativistic treatment became mandatory \cite{reltcc2020}.
In case of the quaternion symmetry, however, the interactions are complex valued, thus the underlying tensor algebra of the DMRG had to be developed for complex numbers as well. 
This affected 
functions related to partial summations to generate the auxiliary operators, the renormalization of the block operators, tasks based on the wave function and the diagonalization of the effective Hamiltonian.
For the complex DMRG version with long range interactions, we also found that the stability of the DMRG performance was more sensitive to accumulated numerical noise, thus numerical roundings have been enforced at several points in the DMRG algorithm.
For the diagonalization step, we further developed the complex version of the Davidson algorithm as well which was found to be very sensitive to even very small numerical noise in the order of $10^{-12}$. 
Alternatively, the L\'anczos algorithm can be used which requires more iteration steps in course of the diagonalization, but provides very stable convergence even in case of small numerical noise in the real and imaginary components. As a hybrid solution, we use the L\'anczos algorithm during the warmup procedure via the dynamic extended active space procedure (DEAS)~\cite{Legeza2003} and the Davidson method for all further DMRG sweeps once initial starting vectors for the diagonalization step is available by the DMRG wave function transformation protocol~\cite{Schollwoeck2011}.
%
Similar developents have been done on the post-DMRG utility functions, i.e., calculation of single orbital entropy, mutual information and obtaining the T1 and T2 amplitudes. The full workflow of the general 4c-TCC in case of quaternion symmetry has been tested and benchmarked against full-CI reference data on small systems.
For larger system sizes, however, the methods for optimization of site ordering turned out to be a delicate issue and procedure from the non-relativistic DMRG turned out in general not to be transferable.
Therefore a particular care has to be taken to the ordering of the sites, which has to keep the A and B spinors from each Kramers pair adjacent.
We used thus the order of adjacent pairs of A,B spinors sorted by HF orbital energy, which is probably not optimal, but the optimization was not able to improve it. Besides the complex-valued arithmetics and larger
number of integrals, this also contributed to the high computational cost of the 4c-DMRG calculations.

Once the CI coefficients  $c_i^a$ and $c_{ij}^{ab}$ have been extracted from the MPS wave function, 
 the standard CC analysis is performed to convert them to the CC amplitudes
\begin{eqnarray}
        \label{eq:ci2cc}
        T^{(1)}_{\text{CAS}} & = & C^{(1)}, \\
        T^{(2)}_{\text{CAS}} & = & C^{(2)} - \frac{1}{2}[C^{(1)}]^2.
\end{eqnarray}
In the present relativistic CC implementation in the Dirac code, one does not distinguish ``spin cases'' of the
excitation amplitudes, but amplitude indices run through the united range of A and B spinors, which transparently includes the excitations with non-zero  $\Delta M_K$. To our advantage, this also matches the single-spinor site representation in the DMRG. In the lowest quaternion-symmetry case, the SD amplitudes 
are in general all non-zero and complex-valued, while in higher symmetry they become sparse, analogously to the
one- and two-electron integrals. Due to the combined effect of doubling the amplitude index range
compared to non-relativistic spin-restricted CCSD and of complex arithmetics required, the low-symmetry relativistic calculations are  computationally more demanding
by a prefactor of almost two orders of magnitude, which restricts the applications to small molecules.

\section{COMPUTATIONAL DETAILS\label{computationalsection}}

\begin{figure}[h]
\begin{tabular}{cc}
\includegraphics[scale=0.125]{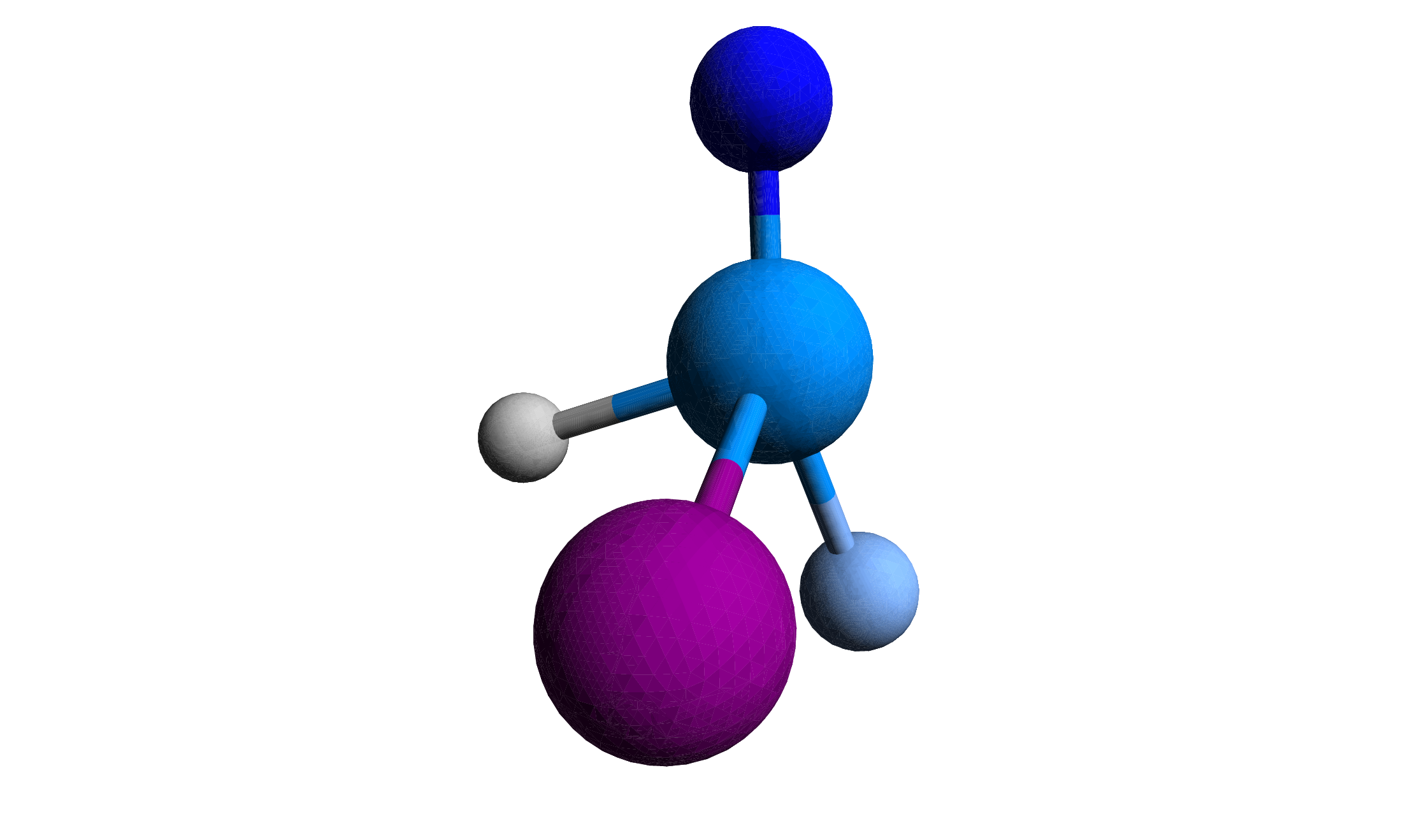} &
\includegraphics[scale=0.125]{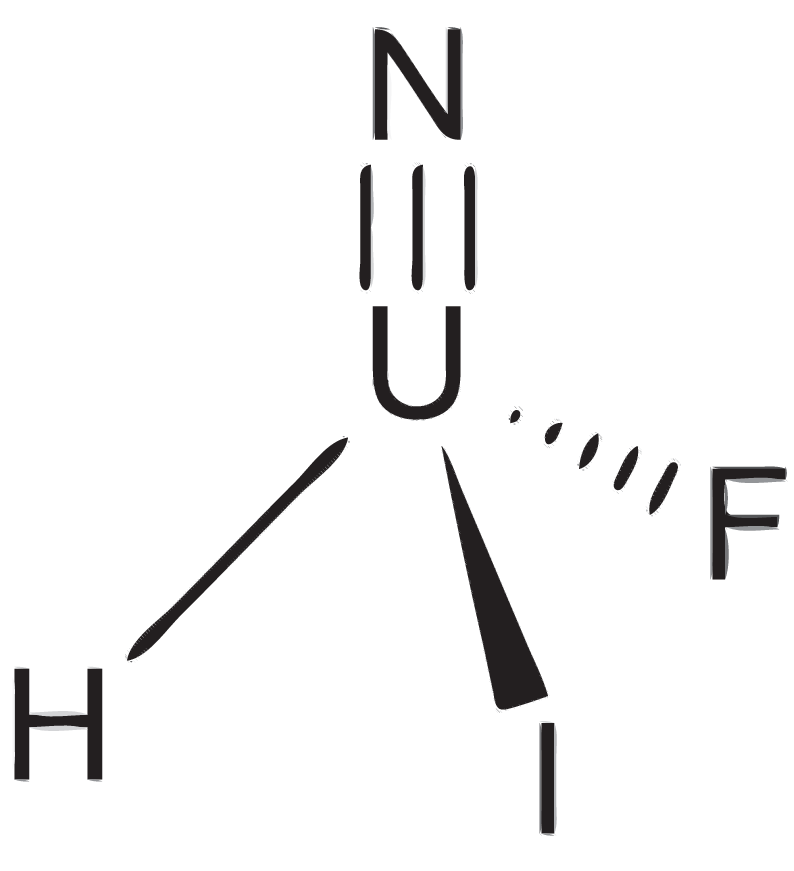} \\ 
\textbf{(S)-NUHFI} & \text{(S)-fluoro(hydrido)}-\\ & \text{iodo(nitrido)uranium}

\end{tabular}
\caption{\label{Fig_xyz} The ECP/B3LYP+D3/def-TZVPP level optimized NUHFI geometry with N-U bond length changed from 1.709 \AA to reference value of 1.700 \AA}
\end{figure}

The molecular geometry of NUHFI (cf. Fig.~\ref{Fig_xyz}) has been optimized at the ECP/B3LYP+D3/def-TZVPP level using Turbomole \cite{TURBOMOLEx}. 

The positions of U, H, F and I atoms have then been fixed and the N atom was moved either closer or more apart from the uranium central atom along the optimized direction of the N-U bond.
The computations along the N-U stretching mode were performed in the four-component relativistic all-electron approach with Dirac-Coulomb Hamiltonian \cite{dyall-faegri} using our development version of the Dirac and Budapest DMRG programs. 
Dyall's double zeta atomic basis sets \cite{dyallbasis} have been used for U and I and non-relativistic contracted cc-pVDZ \cite{ccbasis} for N, H and F (this combination is called "DZ" henceforth).

The selection of the DMRG active space has been based on the single-orbital entropy spectrum calculations in a larger (32,98) space 
with a bond dimension \cite{Schollwoeck2011} up to $M$ = 512 for $R_{NU}$ = 1.4, 1.7, 2.2, and 4.8 \AA\ . 
The notation ($n$,$m$) denotes $n$ correlated electrons in $m$ Kramers pairs, while other electrons are kept inactive.  
Amplitude space for both CCSD and TCCSD (47th to 252nd Kramers pair) 
have been chosen based on the trade-off between computational demands and spectroscopic properties prediction accuracy taking into account energy gaps in the molecular bispinor set. 

For each method considered the UN bond length and force constant were determined by a fit of the UN stretching mode with the programs VIBANAL \cite{vibanal} and TWOFIT associated with DIRAC software. Harmonic frequencies were obtained by using this computed force constant to correct the ECP/B3LYP+D3/def-TZVPP Hessian computed by Turbomole using the relations

\begin{eqnarray}
\frac{\partial R_{UN}}{\partial R_{N_z}} = 1\\
\frac{\partial R_{UN}}{\partial R_{U_z}} = -1\\
H^{corr} = \frac{\partial^2 E^{CC}}{\partial R^2_{UN}} -  \frac{\partial^2 E^{DFT}}{\partial R^2_{N_z}}\\
\frac{\partial^2 E^{CC}}{\partial R^2_{N_z}} =  \frac{\partial^2 E^{DFT}}{\partial R^2_{N_z}} + H^{corr}\\
\frac{\partial^2 E^{CC}}{\partial R^2_{U_z}} = \frac{\partial^2 E^{DFT}}{\partial R^2_{U_z}} + H^{corr} \\
\frac{\partial^2 E^{CC}}{\partial R_{N_z}\partial R_{U_z}} = \frac{\partial^2 E^{DFT}}{\partial R_{N_z}\partial R_{U_z}} - H^{corr}  
\end{eqnarray}

and recomputing the frequencies by rediagonalizing the mass-weighted Hessian. In this mass-weighted Hessian the most abundant isotope mass was selected for each atom.
The first two relations are valid if the bond is aligned along the z-axis, which is a trivial condition to fulfill. 
The other relations require the UN mode to be weakly coupled to the other modes. This assumption was verified by checking the normal modes computed at the DFT level of theory. For both molecules considered, the normal mode considered has less than 2\% admixture of coordinates other than the $R_{N_z}$ coordinate.
The anharmonicities were computed using the program VIBANAL \cite{vibanal} from a polynomial fit of the potential curves along the UN bond employing the effective mass $\frac{1}{\mu } = \frac{1}{m_N}+ \frac{1}{m_U}$.

\section{RESULTS AND DISCUSSION\label{resultssection}}

\begin{figure}[ht]
\centering
\includegraphics[scale=0.5]{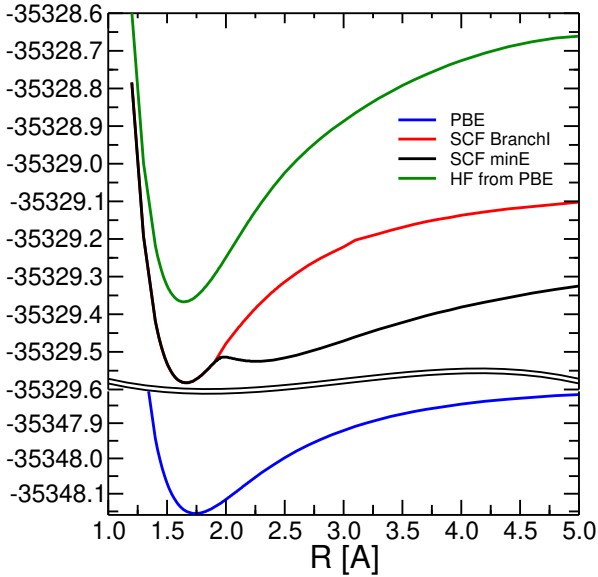}
\caption{\label{Fig.1} SCF and DFT (PBE) potential energy curves}
\end{figure}

The Dirac-Hartree-Fock (DHF) N-U potential energy curve bifurcates at $R$ = 1.90 \AA\   into a lower branch (black line in Fig. ~\ref{Fig.1}) with a double-well character and an upper branch with a shape typical for the ground state of diatomic molecules (red line). The DHF solution of each branch can be followed by restarting the SCF procedure from a converged solution 
at a neighboring geometry. Unfortunately the molecular spinors from either branch lead to unphysical artefacts
in the dissociation curves, when a subsequent DMRG calculation with a limited active space is performed.
The best solution of this problem would probably be a 4c-CASSCF calculation, unfortunately
we did not find any working implementation of this method {\em for quaternion double group symmetry}, neither in Dirac nor in the Bagel and ChronusQ codes (in the version available to us).
Fortunately, the DFT/PBE N-U dissociation curve (blue line in Fig. ~\ref{Fig.1}) exhibits no such instability and bifurcation. Even the single-determinant Dirac-Hartree-Fock energy computed from the self-consistently converged PBE Kohn-Sham molecular spinors (green line in Fig. ~\ref{Fig.1}) is free of the bifurcation or double-well problems.
However, the HOMO-LUMO gap is significantly narrowing after $R$ > 2.0 \AA\ in these orbitals, which leads to convergence problems of CC methods at larger N-U distances. 
There are additional reasons for selecting the PBE functional as a source of molecular spinors: 
We have tried several meta-GGA/hybrid/advanced functionals 
and the Hartree-Fock energy potential energy curves computed from their converged spinors exhibited a non-physical maximum at larger N-U distances (cf. Supplementary material, Fig. S1).
GGA functionals (including PBE) have been found to be free of this artifact and
from the ones we tested the PBE spinors provided the lowest Hartree-Fock energy.
Another reason for the use of PBE Kohn-Sham molecular spinors is the   
lower size of the optimal TCC active space (with respect to minimisation of TCCSD energy at R = 1.700 \AA\  computed in the molecular spinor space 47 \ldots 163). 
For the NUHFI molecule and DHF-SCF spinors this space was (28,40) (at least), while for PBE spinors it is only (20,28) (for further details see Fig.~\ref{Fig.TCCoptim} and Figs. S4 and S5 in the Supplementary Information).
The MP2 natural orbitals yield at R = 1.700 \AA\  an even smaller optimal TCC active space (28,20), but as they are computed starting from the SCF orbitals, they suffer from the bifurcation problem as well.
Finally, we found the PBE functional adequate to describe other molecules containing the U-N bond (see Appendix). 
We have thus based our post-HF calculations on the PBE Kohn-Sham molecular spinors.

\begin{figure}[ht]
\centering
\includegraphics[scale=0.5]{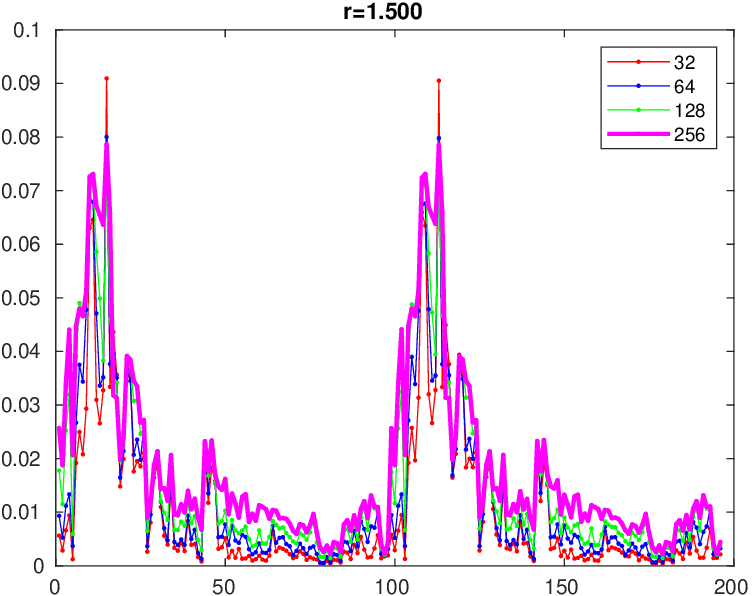}
\caption{\label{Fig.entropy}DMRG(32,98) entropy profiles at $r_\mathrm{N-U}=1.5$\AA$\;$for various bond dimensions}
\end{figure}

In order to select the DMRG active space we performed preliminary DMRG calculations
with a very large active space (32,98) at lower bond dimensions and for several
values of the bond length. Fig.~\ref{Fig.entropy} shows the results for $r_\mathrm {N-U}=1.5$\AA, while the results for other distances can be found in the supplementary material.  
The results for A and B spinors form two almost identical blocks and the bond dimensions 128 and 256 yield very similar results, indicating sufficient accuracy for this purpose. Based on the entropy profiles, we have chosen three active spaces, small (14,15), medium (24,26), and large (32,32). 
The orbital entropies yield a quantitative indication how important individual orbitals are and
within the DMRG method the more one can afford to include in the active space, the better. However, in the TCC method, a too large active space might not be ideal,
due to the ``freezing'' of the static correlation in this space, which does not
interact with the dynamical correlation effects outside the active space. As shown previously \cite{Faulstich2019b,reltcc2020}, in the TCC method a smaller active space may thus yield better results than a bigger one. We have thus also performed an attempt to find the optimum active space for TCCSD, cf. Fig.~\ref{Fig.TCCoptim}, using the DMRG bond dimension $M=1024$
and CC orbital space restricted to 47 \ldots 163.
As can be seen, the space (20,28) seems to perform best in terms of yielding lowest TCC energy at the bond distance 1.7 \AA. Of course, at a different bond distance the outcome might be different and it is clearly not practical to perform such optimization globally.
We have thus confined the detailed calculation of the potential energy curves to the three entropy-selected active spaces and this ``TCC-optimal'' active space.

\begin{figure}[ht]
\centering
\includegraphics[scale=0.35]{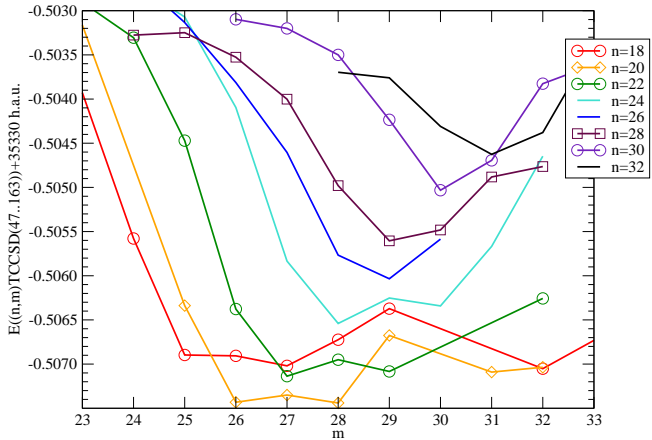}
\caption{\label{Fig.TCCoptim}Active space optimization for DMRG($n$,$m$)-TCCSD single point computation of NUHFI molecule (R = 1.700 \AA) in PBE/DZ basis set, bond dimension $M=1024$, CC orbital space 47\ldots 163}
\end{figure}

Besides the entropies, the DMRG(32,98) calculations yielded potential energy curves
(at a very coarse bond length grid), which are given in Fig.~\ref{Fig.dmrg32,98}.
It can be seen that at the bond dimensions affordable for this very large space, the energies are far from saturation. Since the computational cost of these calculations
with large enough bond length would be intractable, we did not attempt to refine the bond length grid here, but performed further DMRG and TCC calculations employing other
smaller active spaces selected using either the orbital entropy criteria (Fig.~\ref{Fig.entropy}) or the minimum TCCSD energy at R = 1.70 \AA (Fig.~\ref{Fig.TCCoptim}).

\begin{figure}[ht]
\centering
\includegraphics[scale=0.5]{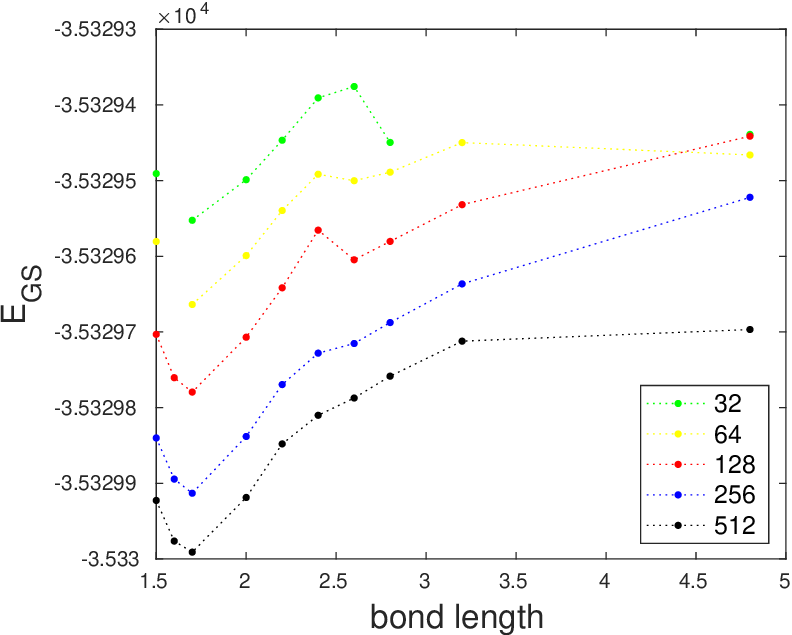}
\caption{\label{Fig.dmrg32,98}DMRG(32,98) potential energy curves for various bond dimensions}
\end{figure}

\begin{figure}[ht]
\centering
\includegraphics[scale=0.30]{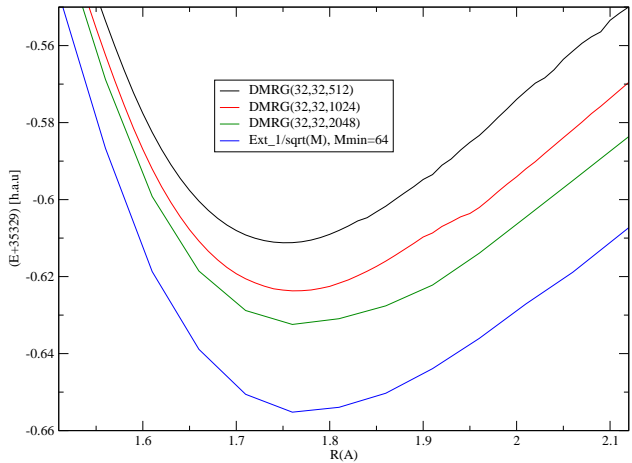} 
\caption{\label{Fig.MiniP2} N$\equiv$U potential energy curves for DMRG(32,32,M)}
\end{figure}

We have also investigated the effect of the bond dimension on the DMRG potential energy curves in the largest selected active space (32,32), which are shown in Fig.~\ref{Fig.MiniP2}. 
Here  the computational limitation was $M=2048$ and in spite of this
relatively large bond dimension, the DMRG does not seem close to saturation either. We extrapolated DMRG(32,32,$M$) energies to infinite bond dimension $M$ using a simple
second order polynomial and a 
three-parameter model
\begin{eqnarray}
\label{extrapolf}
E(M) = E(M\to +\infty)+\frac{a}{M^p},
\end{eqnarray}
where $E(M)$ is the DMRG(32,32,$M$) energy, $E(M\to +\infty)$ its extrapolated value and $a$ and $p$ are positive fit parameters. 
E(M) data have been collected from $M$ = 64, 128, 256, 350, 512, 750, 1024, 2048. (Of course, the bond dimension has a finite maximum value in any finite AO basis, but a large enough to make such an extrapolation sensible.)
We have found that the best fit with lowest
residual error has been obtained by fixing
$p=-1/2$ which is usually applied to fit
energy gaps in
gapped systems (to which molecules belong), where the correlations decay exponentially. 
We note that the extrapolation using a general power $p$ of (\ref{extrapolf}) did not lead to significantly different values and was numerically unstable for R > 1.86 \AA.
The fit via the second order polynomial was very unreliable. 
A more rigorous fit based on the DMRG truncation error could not be applied here, due to very limit range of the available truncation error data points.
The associated spectroscopic constants are shown in  Tab.~\ref{spectroscopic2}.

\subsection{Assessment of TCC in a small orbital space with respect to extrapolated DMRG as a benchmark}

As a first step we decided to compare the performance of TCC with respect to
DMRG extrapolated to infinite bond dimension, within a small virtual orbital subspace. We have chosen DMRG(32,32) space, which corresponds to the range of 66th-97th Kramers pairs in the CC orbital space.
We performed 4c-TCC in several subspaces, namely (14,15)TCCSD, (20,26)TCCSD, (28,29)TCCSD and (24,26)TCCSD,
as well as CCSD and CCSD(T) for comparison.
The DMRG(32,32) has been extrapolated to the $M \to \infty$  limit 
according to (\ref{extrapolf})
which should be equivalent to CASCI(32,32). 

\begin{figure}[ht]
\centering
\includegraphics[scale=0.35]{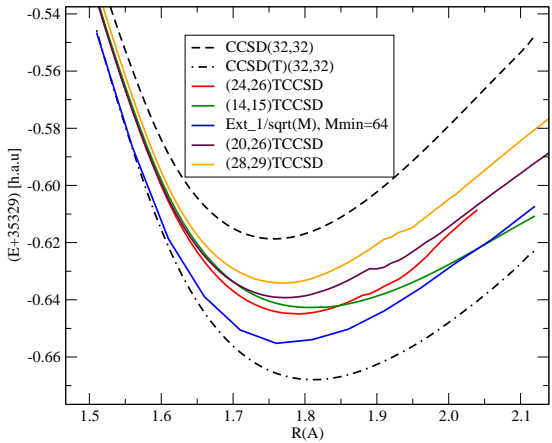}
\caption{\label{Fig.MiniP1} N$\equiv$U potential energy curves computed with methods restricted to the (32,32) orbital space}
\end{figure}

As can be seen from Fig.~\ref{Fig.MiniP1}, the order of energies from different methods is
\begin{eqnarray}
\label{order}
\rm{CCSD(T)}  <  \rm{DMRG}(M \to  \infty)  < \rm{TCCSD } <  \rm{CCSD}, 
\end{eqnarray}
where all CC methods have amplitude range confined to the DMRG(32,32) subspace.
Interestingly, the CCSD(T) curve lies below the extrapolated DMRG benchmark,
indicating that the perturbative triples correction overshoots 
the exact result or the DMRG extrapolation is not able to reproduce the exact (FCI) energy accurately enough.
From the potential energy curves in Fig.~\ref{Fig.MiniP1}, spectroscopic constants have been computed and are listed in Tab.~\ref{spectroscopic2}. 
The results indicate that (24,26)TCCSD provides the closest minimum energy to the extrapolated DMRG(M=+$\infty$).
 The vibrational frequency for (24,26)TCCSD agrees well with CCSD(T) while being about 70cm$^{-1}$ below the extrapolated DMRG. The (24,26)TCCSD  anharmonicity 
 is close to the exptrapolated DMRG, while standard CCSD and CCSD(T) yield lower anharmonicity values.
Tailored CC spectroscopic constants are rather sensitive to the active space chosen, which stresses the neccesity of appropriate active-space choice for TCCSD. 
Inside the orbital space (32,32) used as a restriction for both amplitudes and active orbitals, it is not possible to go reasonably much above (24,26), as there will hardly be any amplitudes left. Performing TCC in such a small
virtual orbital space, although useful as a sanity check, clearly is not in the spirit of using TCC to capture the dynamic correlation.
We attempted a similar test in a larger space corresponding to DMRG(32,98),  which was the space used for orbital entropy calculations at low $M$, but it turned out to be computationally too costly. 


\begin{table}[ht]
\caption{\label{spectroscopic2} Spectroscopic parameters for the N-U bond of the NUHFI molecule in the Dyall-v2z(U,I) and cc-pVDZ (N,H,F) basis, restricted to the (32,32) orbital space}
\begin{tabular}{|l|l|l|r|r|}
\hline
 Method        	&  $r_{\rm{e}}$  &   $E_{\rm{min}}$ & $\omega_{\rm{e}}$ & $\omega_{\rm{e}}\rm{x}_{\rm{e}}$  	\\ 
  	&  \AA &  a.u. & $\rm{cm}^{-1}$ &  $\rm{cm}^{-1}$ 	\\
 \hline
  DMRG(M=1024,d)\footnotemark[3]{}   	& 1.7591 & -35329.62361 & 1062.42 & 17.19 \\ 
  DMRG(M=1024,s)\footnotemark[4]{}   	& 1.7567 & -35329.62348 & 1053.08 & 19.50 \\ 
  DMRG(M=2048,s)\footnotemark[4]{} & 1.7727 & -35329.63243 & 946.67 &  8.92 \\ 
  DMRG(M=+$\infty$,s)\footnotemark[4]{}\footnotemark[5]{} & 1.7766 & -35329.65525  & 1006.63 & 10.04   \\ 
  CCSD    &  1.7561  & -35329.61877 & 1084.92 & 5.44   \\    
 CCSD(T)	   	& 1.8110  & -35329.66792 &   929.38 &  3.17 \\ 
 (14,15)TCCSD  	& 1.8088 & -35329.64278 & 812.16 &  2.80     \\ 
 (20,26)TCCSD  &   1.7680 & -35329.63902 &    998.56  & 22.80 \\ 
 (24,26)TCCSD   & 1.7852 & -35329.64465 & 935.77 &  10.91 \\ 
 (28,29)TCCSD &   1.7637 & -35329.63398 &   1042.37 &  11.52 \\ 
    \hline
\end{tabular}
\footnotetext[3]{Dense sampling of distances, with dR = 0.01 \AA  step. That have been used for all CC methods as well.}
\footnotetext[4]{Spare sampling of distances, with dR = 0.05 \AA  step}
\footnotetext[5]{Extrapolation to infinity M done with respect to $E = E(M\to+\infty)+a/\sqrt{M}$ formula and with set starting with $M_{min} = 64$ and all powers of 2 to $M_{max} = 2048$ together with $M = 350$ and $M = 750$}
\end{table}



\subsection{Potential energy curves and spectroscopic constants}

For the CC and TCC calculations we have chosen the orbital space comprised by spinors 47 \ldots 252, as a compromise between accuracy and computational demands. Although it is a severe truncation of the total orbital space in the employed basis set, it already provides a substantial number of external orbitals not in the DMRG active space to allow TCC to capture a significant portion of the dynamic correlation.   

\begin{figure}[ht]
\centering
\includegraphics[scale=0.5]{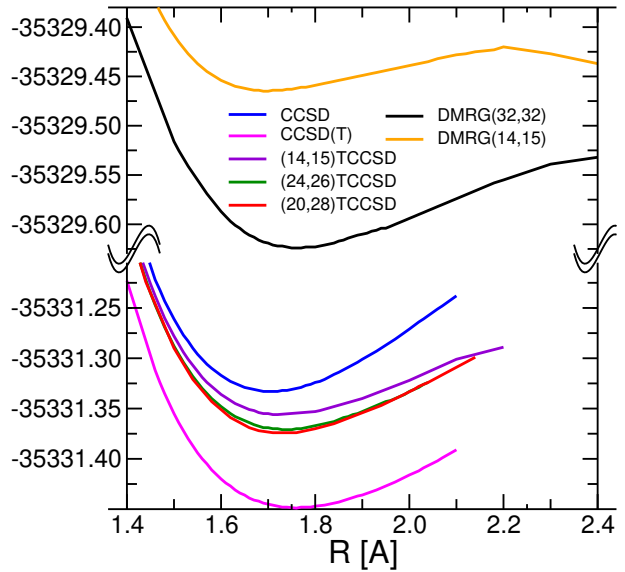}
\caption{\label{Fig.2}DMRG and CC N$\equiv$U potential energy curves. The range of amplitudes have been 47..252 for all CC computations}
\end{figure}

\begin{figure}[ht]
\centering
\includegraphics[scale=0.35]{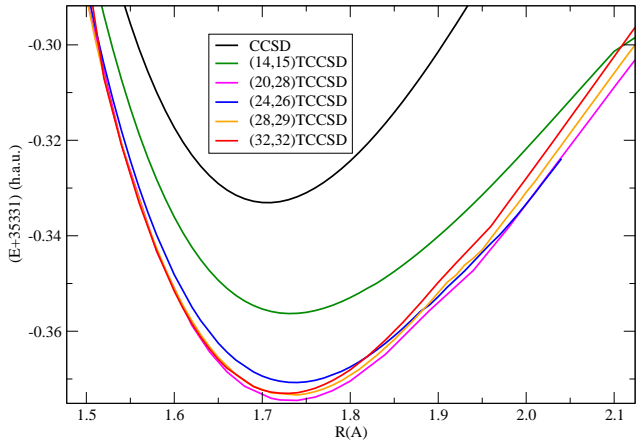}
\caption{\label{Fig.DZ252TCCselect}CC and TCC potential energy curves for various active space sizes, amplitude range from 47th to 252th PBE Kramers pair}
\end{figure}

The potential energy curves of NUHFI computed by DMRG, CC, and TCC methods are plotted in Fig. ~\ref{Fig.2}, while a detailed comparison of TCC results is presented in Fig.~\ref{Fig.DZ252TCCselect} at a finer energy scale.
Notice that the TCCSD curves always lie between standard single-reference CCSD and CCSD(T) ones. 
The CC amplitude equations in TCCSD also converge faster and for larger N-U distances, where ordinary CCSD/CCSD(T) already fails. 
The better convergence of the tailored CCSD can be explained by two effects: (i) the most problematic amplitudes involving close-lying orbitals around HOMO and LUMO (responsible for the static correlation) are determined by DMRG and fixed and (ii)  there are less amplitudes remaining to be optimized.

\begin{table}[ht]
\caption{\label{spectroscopic} Spectroscopic parameters for the N-U bond of the NUHFI molecule in the Dyall-v2z(U,I) and cc-pVDZ (N,H,F) basis}
\begin{tabular}{|l|l|l|r|r|}
\hline
 Method        	&  $r_{\rm{e}}$  &   $E_{\rm{min}}$ & $\omega_{\rm{e}}$ & $\omega_{\rm{e}}\rm{x}_{\rm{e}}$  	\\ 
  	&  \AA &  a.u. & $\rm{cm}^{-1}$ &  $\rm{cm}^{-1}$ 	\\
 \hline
 4c-DHF-SCF	   	& 1.6614  & -35329.58205    & 1280.23       &    3.59   	            			\\ 
 4c-PBE	   	&  1.7408 & -35348.15605    & 1018.67
   &   4.43     	    			\\ 
  4c-DHF(PBE)	   	& 1.6437  & -35329.36686    & 1338.49       &   3.08       			\\ 
  \hline
  4c-B3LYP	   	& 1.7224  & -35346.76292   &   1075.25 	  &    3.47		\\ 
 ECP/B3LYP,TZ\footnotemark[1]{}&	  & 		  &  1035.0	  &                       \\ 
 4c-B3LYP,TZ\footnotemark[3]{}&	  & 		  &  1065.4	  &                       \\ 
 \hline
 4c-DMRG(14,15)\footnotemark[8]{} &      1.7016 &  -35329.46503 &  904.28 &  17.65 \\ 
 4c-DMRG(20,28)\footnotemark[9]{} & 1.7674 & -35329.58429 &   1001.29 &  4.81 \\ 
4c-DMRG(24,26)\footnotemark[10]{}      &    1.7547 & -35329.54655 &    847.93 &  3.78 \\ 
4c-DMRG(28,29)\footnotemark[11]{} &    1.7797 & -35329.59360  &  982.9 &  9.92 \\ 
 4c-DMRG(32,32)     &   1.7634 & -35329.62372 &    1060.59 & 7.14    \\ 
 \hline
 4c-CCSD\footnotemark[6]	   	& 1.7059  & -35331.33308    &  1139.34     &    3.29		\\ 
 4c-CCSD(T)\footnotemark[6]	   	& 1.7604  & -35331.44856    &   973.93     &    2.99	\\ 
 \hline
 4c-(14,15)TCCSD\footnotemark[6]  	& 1.7324  & -35331.35627    & 969.2 & 7.63  \\ 
 %
 \textbf{4c-(20,28)TCCSD}\footnotemark[6] \footnotemark[7] & \textbf{1.7314} & \textbf{-35331.37442} &   \textbf{1071.77} & \textbf{7.24} \\ 

 4c-(24,26)TCCSD\footnotemark[2]{} \footnotemark[6] & 1.7353 & -35331.06927  &  1036.53 & 6.62 \\  %
 4c-(28,29)TCCSD\footnotemark[4]{} \footnotemark[6]\footnotemark[12]{} &       1.7308 & -35331.37340 &   1076.93 & 6.74 \\ 
 4c-(32,32)TCCSD\footnotemark[6]  	& 1.7250  & -35331.37310    & 1104.49     &   8.17	\\ 
 \hline
 ECP\footnotemark[5]/CCSD(F12) & 1.6819 & -642.23964 & 1149.47 & 3.58 \\ 
 ECP\footnotemark[5]/CCSD(T)(F12)\footnotemark[2] & 1.7061 &  -642.30227    &  1095.43 & 2.69 \\ 
 ECP\footnotemark[5]/CCSD(T*)(F12)\footnotemark[2] &  1.7079 &  -642.31204  & 1092.03 & 2.45  \\ 
 \hline
 ECP\footnotemark[5]/DFT:r2SCAN & 1.7114 &  -643.73867 &      1061.35 & 3.38  \\ 
 ECP\footnotemark[5]/DFT:mpsts-noa2 & 1.7056 & -643.51394  & 1077.54 & 3.34 \\ 
    \hline
\end{tabular}
\footnotetext[1]{Scalar quasi-relativistic computation From Ref.\cite{Wormit2014}}
\footnotetext[3]{Computation From Ref.\cite{Wormit2014}, harmonic value, four-component Hamiltonian}
\footnotetext[2]{R$_{max}$ = 2.040 \AA}
\footnotetext[4]{Sampled with half frequency, dR = 0.02 \AA}
\footnotetext[5]{Scalar quasi-relativistic, with def2-TZVPP AO basis, computed using Turbomole V7.6\cite{TURBOMOLEx},\cite{TURBOMOLEx2}}
\footnotetext[6]{4c-CC and almost all 4c-TCC computations have been done with amplitude range 47..252}
\footnotetext[7]{The bold text corresponds to the optimal active space for the 4c-TCCSD method selected by minimisation of 4c-TCCSD energy at R = 1.700 \AA\ (See text for more details.)}
\footnotetext[8]{Bond dimension M = 2048, number of sweeps = 41}
\footnotetext[9]{R$_{max}$ = 1.850 \AA}
\footnotetext[10]{R$_{max}$ = 1.880 \AA}
\footnotetext[11]{R$_{max}$ = 1.900 \AA}
\end{table}

Spectroscopic constants in Table~\ref{spectroscopic} have been computed as specified in the computational details.
The part of dissociation curve delimited by a maximal N-U distance $R_{max}$ = 2.100 \AA $ $ and a minimal $R_{min}$ corresponding to the same energy has been used for the fit, 
unless indicated otherwise in the footnotes.

As there are no experimental data for NUHFI molecule in so far, only comparison with another computational study \cite{Wormit2014} has been possible. 

Table~\ref{spectroscopic} summarizes our results computed by various methods, as well as data from the available literature. 
The DMRG and 4c-TCCSD results are presented for several
DMRG active spaces, which have been chosen base on orbital entropies, and also optimized as described in the previous section. The results for the optimal active space (20,28) have been indicated by boldface.

Compared to all correlated methods, the 4c-DHF-SCF underestimates the bond length and anharmonicity, while it overestimates the vibrational frequency. DHF computed from the PBE orbitals (employed as a reference for post-HF methods in this work) shows an even stronger tendency in this direction, unsurprisingly demonstrating the importance of the correlation effects.

We performed also a 4c-B3LYP calculation in the basis set employed in this work and compared it to the B3LYP vibrational frequencies in a larger basis set from Ref.~\cite{Wormit2014}.
There is no dramatic effect of the basis set size on the vibrational frequency, which supports the use of our basis set of choice. (Since we had to truncate the virtual orbital space for CC calculations, the use of a larger AO basis for our calculations would not be sensible anyway.)

Next part of Table~\ref{spectroscopic} shows DMRG results in several active spaces. In the smallest one (14,15), the bond length is rather close to the DHF value, while it becomes larger 
for the larger spaces. The DMRG vibrational frequencies lie below 1000 cm$^{-1}$ except for the (20,28) and (32,32) spaces, while the anharmonicities differ over a wide range. Particularly the anharmonicity from smallest DMRG space (14,15) is an outlier. 
The reason might be that these spaces are rather small and addition of a few more orbitals has thus a large effect. 
However, as can be seen on next lines of the table, the 4c-TCCSD method, although based on DMRG, yields anharmonicity values in a rather narrow range,
even for the small space, where DMRG value was extremely large.

The 4c-TCCSD method yields similar values of the equilibrium distance (around 1.73 \AA) regardless of the active space. 
The vibrational frequency is more sensitive on the active space choice, but except
for the smallest space (14,15), it always lies above 1000cm$^{-1}$.
Anharmonicities are again less sensitive on the active space, being all around 7 cm$^{-1}$.
Compared to standard single-reference 4c-CCSD and 4c-CCSD(T), the 4c-TCCSD equilibrium distance and vibrational frequency lies between the 4c-CCSD and 4c-CCSD(T) values. This might be explained by the fact that these properties are determined by the PEC around equilibrium, where the
multireference character is still weak and 4c-TCCSD can capture more
of the dynamic correlation than 4c-CCSD but less than 4c-CCSD(T).
On the other hand, the 4c-TCCSD anharmonicity is approximately twice as large than the 4c-CCSD or 4c-CCSD(T) values, which are similar.
Since already the DMRG method alone yielded larger anharmonicities,
although widely varying in dependence on the active space, one
can deduce that the strong correlation is more important for
the anharmonicity, in line with the fact that the multireference character of the system grows with the elongation of the N-U bond.
Single-reference methods like CCSD or CCSD(T) thus probably underestimate the anharmonicity. It is possible that DMRG or 4c-TCCSD might on the other hand overestimate it, but for the NUHFI molecule we unfortunately do not have an experimental reference.
We thus performed also calculations on the NUF$_3$ molecule, where experimental $\omega_e$ and $\omega_ex_e$
data are available, as will be described in the next section.

Finally, we performed also scalar ECP quasirelativistic calculations
with CCSD and CCSD(T) in their explicitly correlated F12 versions to better capture the dynamic correlation and advanced DFT methods (r2SCAN is regularized-restored Strongly Constrained and Appropriately Normed meta-GGA functional \cite{r2scan}, mpsts-noa2 is (modified) functional of Perdew, Staroverov, Tao, and Scuseria (PSTS-LMF). \cite{mpsts-noa2a},\cite{mpsts-noa2b} ) 
implemented in the Turbomole code\cite{TURBOMOLEx},\cite{TURBOMOLEx2}, using a larger def2-TZVPP basis set.
The bifurcation of the SCF solution in this case occurred at a larger bond length,
so we were able to obtain smooth curves for the post-HF methods in a large enough range to compute the spectroscopic constants.
All these methods yield slightly shorter bond length than the 4c-TCCSD methods, which might be due to the 
different basis or due to different treatment of the relativity, but
the vibrational frequencies at the CCSD(T)-F12 level are 
close to the 4c-TCCSD results discussed above. 
The DFT methods yield actually very similar results to 4c-CCSD(T). 
Again, all these methods yield anharmonicities smaller than 4c-TCCSD, which is probably due to their single-reference character.

\subsection{Comparison with the NUF$_{3}$ molecule}

Due to the lack of experimental spectroscopic data for NUHFI, we also investigated a similar molecule, NUF$_{3}$, for which the vibrational frequency of the N-U stretch has been measured\cite{Andrews2008}.
The experimental anharmonicity is still not available, so we at least tried to
make a plausible estimate  based on the measured anharmonicity 5.3 cm$^{-1}$ in NU$^+$ \cite{VanGundy_phd} with an error bar of $\pm 8\%$, 
which corresponds to the spread of the N-U bond vibrational frequencies 
in the set of related species (UN$^{+}$\cite{VanGundy_phd}, UN\cite{KING20142}, NUO$^{+}$\cite{KING20142}, NUN\cite{KING20142}, NUF$_{3}$\cite{KING20142},\cite{Atkinson2018}, NUNH\cite{KING20142})

\begin{figure}[ht]
\centering
\includegraphics[scale=0.25]{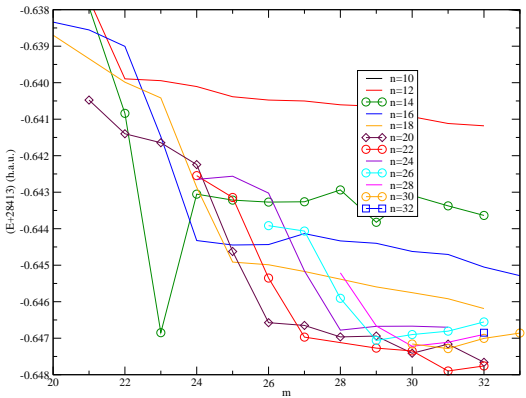}   
\caption{\label{Fig.TCCoptimNUF3}Active space optimization for ($n$,$m$)-4c-TCCSD single point computation of NUF$_3$ molecule (R = 1.740 \AA) in PBE/DZ basis set}
\end{figure}

The optimal active space $(n,m)$ for 4c-TCCSD computations of the N-U bond stretching mode in NUF$_{3}$ molecule has been evaluated by an analogous procedure as in the NUHFI molecule. We minimized the energy E(($n,m$)4c-TCCSD(35 \ldots 172)) with respect to $n$ and $m$ (cf. Fig.~\ref{Fig.TCCoptimNUF3}). 
The lowest energy has been achieved for the space of 22 electrons in 31 orbitals (22,31), followed by (20,32). 
The orbital range in 4c-CCSD, 4c-CCSD(T) and 4c-TCCSD calculations for NUF$_{3}$ has been always 35..172 and will be henceforth omitted.

Table~\ref{spectroscopic3} presents resulting spectroscopic constants of NUF$_3$. Similarly to the NUHFI molecule, uncorrelated results from DHF-SCF and DHF from PBE spinors strongly underestimate the bond length and overestimate the vibrational frequency. The bifurcation of the DHF-SCF solution occurs at R = 1.93 \AA, as can be seem from Fig.~S2 in the Supplementary information. 
The PBE functional, which has been employed to obtain molecular spinors for subsequent post-HF calculations due to its bifurcation-free behavior, yields a bond length in a good agreement with CASPT2(6,16) calculations \cite{Atkinson2018} and vibrational frequency reasonably close to the experimental value \cite{Andrews2008}.

Concerning the single-reference CC methods, addition of perturbative triples substantially elongates the bond and lowers the vibrational frequency, exemplifying the importance of the dynamic correlation.
DMRG yields consistently bond lengths close to the CCSD(T) value, while the vibrational frequencies are too small compared to the experimental value,
except in the largest (32,32) space, where $\omega_e$ is close to the experiment. However, DMRG in this space yields a too long bond length and too high anharmonicity.

Similarly to NUHFI, DMRG yields widely varying anharmonicities for different active spaces.
However, in constrast to NUHFI, here the DMRG in the smallest active space yields the smallest anharmonicity. 


The 4c-TCCSD method yields similar bond lengths in all active spaces, which are slightly shorter than 4c-CCSD(T) value and close to the CASPT2(6,16) value from the literature \cite{Atkinson2018}. The vibrational frequencies are in the range of 920-1010 cm$^{-1}$, the one for the optimal space (22,31) being 977 cm$^{-1}$, which is reasonably close to the experimental value of 938 cm$^{-1}$. The anharmonicities are in this case more sensitive to the active space choice than in the NUHFI molecule. 
Small active spaces yield values close to the single reference methods, while larger spaces yield larger anharmonicity values. 
For the optimal space (22,31), the result is 6 cm$^{-1}$, which is in line with the estimate based on the experimental data of similar molecules \cite{VanGundy_phd}. 

We have also performed calculations employing scalar ECP quasirelativistic CC methods with explicit correlation and in a larger basis set. 
The ECP-HF-SCF solution has again a bifurcation at 1.93 \AA and the post-HF results depend on which SCF branch has been chosen as reference (see footnotes in Table~\ref{spectroscopic3}).
All versions of the CCSD(T) method in this case yield somewhat shorter bond length than the CCSD(T) in the full 4-component treatment, while the vibrational frequencies are consistently too high.
The anharmonicity depends on whether lower or upper branch of the bifurcated SCF solution has been employed. The upper branch, which does not have an unphysical maximum in the SCF energy, is probably a better choice and yields CCSD(T)-F12 anharmonicities around 5 cm$^{-1}$, which is close to the estimate.
The ECP/DFT results yield similar bond lengths as the CC methods, their vibrational frequencies being substantially over the experimental one and anharmonicites below the estimate.
It thus seems that a combined description of the dynamic and static correlation is needed for this molecule as well in order to describe its spectroscopic properties, particularly 
that DMRG in modest spaces is not able to yield reliable anharmonicities.


\begin{table}[ht]
\caption{\label{spectroscopic3} Spectroscopic parameters for the N-U bond of the NUF$_3$ molecule in the Dyall-v2z(U) and cc-pVDZ (N,F) basis}
\begin{tabular}{|l|l|l|r|r|}
\hline
 Method        	&  $r_{\rm{e}}$  &   $E_{\rm{min}}$ & $\omega_{\rm{e}}$ & $\omega_{\rm{e}}\rm{x}_{\rm{e}}$  	\\ 
  	&  \AA &  a.u. & $\rm{cm}^{-1}$ &  $\rm{cm}^{-1}$ 	\\
 \hline
4c-DHF-SCF\footnotemark[16]{} & 1.6704 & -28412.17024 &  1282.86 & 5.64 \\ 
%
4c-PBE & 1.7567 & -28428.26886 &   973.23 &  3.88 \\ 
4c-DHF(PBE) & 1.6553 & -28411.96614 &   1307.93 & 2.98 \\ 
\hline
4c-CCSD\footnotemark[4]{} & 1.7212 & -28413.59978 &    1106.06 & 3.06 \\ 
%
4c-CCSD(T)\footnotemark[4]{} & 1.7787 & -28413.70853 &   930.2 &  2.74 \\  
\hline
4c-DMRG(14,15) & 1.7565 & -28412.08719 &   805.61 & 3.25 \\ 
4c-DMRG(18,18) &   1.7714 & -28412.10742 &    827.73 &  5.76 \\ 
4c-DMRG(26,26)\footnotemark[8]{} &        1.7777 & -28412.16795 &    821.37 &  6.89 \\ 
4c-DMRG(22,31) & 1.7904 & -28412.19696 &  873.91 &   4.60  \\ 
4c-DMRG(32,32) & 1.7908 & -28412.23037 &   910.19 &  8.81 \\ 

\hline
4c-(14,15)TCCSD\footnotemark[4]{} & 1.7661 & -28413.63386 &   923.95 &  2.83 \\ 
4c-(18,18)TCCSD\footnotemark[4]{} & 1.7659  & -28413.63386 &    935.74 &  3.53    \\ 
4c-(20,32)TCCSD\footnotemark[4]{} &  1.7576 & -28413.64760 &   964.93 &  4.47 \\ 
\textbf{4c-(22,31)TCCSD}\footnotemark[4]{} \footnotemark[11]{} & \textbf{1.7547} & -28413.64762  &  \textbf{977.19} &  \textbf{5.99} \\           
4c-(26,26)TCCSD\footnotemark[4]{} & 1.7587 & -28413.64416  &  954.6 &  5.36      \\ 
4c-(32,32)TCCSD\footnotemark[4]{} & 1.7493 & -28413.64722 & 1008.88 & 6.89 \\ 
   \hline
ECP\footnotemark[5]{}/CCSD(F12) & 1.6995 & -829.92403  & 1104.63 & 5.34   \\ 
ECP\footnotemark[5]{}/CCSD(T)(F12)\footnotemark[7]{}  &    1.7218  & -829.99853 &    1061.62 & 2.80  \\ 
ECP\footnotemark[5]{}/CCSD(T)(F12)\footnotemark[10]{} & 1.7200 & -829.99844 &  1058.26 & 4.95  \\ 
ECP\footnotemark[5]{}/CCSD(T*)(F12)\footnotemark[7]{} &  1.7247 & -830.00965  & 1083.8 & 5.07 \\ 
ECP\footnotemark[5]{}/CCSD(T*)(F12)\footnotemark[10]{} & 1.7216 & -830.00941 & 1053.39 & 5.59  \\ 
\hline
ECP\footnotemark[5]{}/DFT:r2SCAN & 1.7093 & -831.52519  &    1000.59 & 3.43 \\ 
ECP\footnotemark[5]{}/DFT:mpsts-noa2  & 1.7279 & -831.33322  &    1019.11 & 3.50 \\ 
\hline
CASPT2(6,16)\footnotemark[12]{} & 1.753 & & & \\
\textbf{Exp.}\footnotemark[13]{} &   &  & \textbf{938}	&  \\
Estimate\footnotemark[14]{} &&&& 5.3 $\pm$ 0.4 \\
\hline
\end{tabular}
\footnotetext[4]{All 4c-CC and 4c-TCC computations have been done with amplitude range 35..172}
\footnotetext[5]{Scalar quasi-relativistic, with def2-TZVPP AO basis, computed via Turbomole V7.6\cite{TURBOMOLEx},\cite{TURBOMOLEx2}}
\footnotetext[6]{The curve is limited by $R_{max}$ = 2.02 \AA}
\footnotetext[7]{The curve is limited by $R_{max}$ = 2.00 \AA}
\footnotetext[8]{The curve is limited by $R_{max}$ = 1.90 \AA}
\footnotetext[9]{The curve is limited only by $R_{max}$ = 2.10 \AA\ but higher branch of the bifurcated HF-SCF solution was taken for longer distances of R > 1.95 \AA}
\footnotetext[10]{The curve is limited only by $R_{max}$ = 2.10 \AA\ but higher branch of the bifurcated HF-SCF solution was taken for longer distances of R > 1.93 \AA}
\footnotetext[11]{The bold text corresponds to the optimal active space for the 4c-TCCSD method selected by minimisation of 4c-TCCSD energy with X = 163 virtual orbital cut-off at R = 1.700 \AA\ distance.}
\footnotetext[12]{From Ref.~\cite{Atkinson2018}}
\footnotetext[13]{Experimental data from Ref.~\cite{Andrews2008} indicated in boldface}
\footnotetext[14]{Estimate based on the experimental anharmonicity value for NU$^+$ \cite{VanGundy_phd} and a spread of vibrational frequencies in a group of similar molecules}
\end{table}



\clearpage

\section{CONCLUSIONS}
We have implemented the relativistic tailored coupled clusters method (4c-TCCSD), in combination of the Budapest DMRG and Dirac codes.
The method is capable of treating relativistic, strongly correlated 
systems and include a significant portion of the dynamical correlation.
The present implementation is fully general, applicable to polyatomic molecules and is not limited to a particular molecular symmetry double group.

We have applied the method to the study of the stretch of the nitrogen-uranium bond in the NUHFI and NUF$_3$ molecules. This bond can be viewed as a relativistic analogue of the triple bond in N$_2$, which is well known for
a multireference character at stretched bond lengths.
The chiral NUHFI molecule is also interesting as a candidate for measurable electroweak parity violating effects, while for NUF$_3$ experimental vibrational
frequency is available.

It turned out that the choice of the molecular spinors (orbitals) for the post-HF calculations of these molecules is non-trivial, as the DHF-SCF solutions for both molecules exhibit a bifurcation to two branches around 2\AA\ . 
Relativistic CASSCF (or DMRG-SCF) calculation was not possible for the
low-symmetry NUHFI molecule with presently available codes, so we have employed
PBE Kohn-Sham spinors as the molecular spinor basis.

The 4c-TCCSD method with optimal active space yielded for NUF$_3$ vibrational frequency about 40 cm$^{-1}$ above the experimental value, while the anharmonicity was in line with an estimate based on experimental value for similar molecules. This agreement was considerably better than the performance of DMRG or single-reference 4c-CCSD alone.
For both molecules, the 4c-TCCSD method yielded bond lenghts and vibrational frequencies between the values from single-reference 4c-CCSD and 4c-CCSD(T) methods, indicating that 4c-TCCSD is able to capture more dynamical correlation than 4c-CCSD,
but less than 4c-CCSD(T). 4c-TCCSD yielded larger anharmonicity values than the single-reference CC methods, which can be attributed to the better description
of the potential energy curves at larger distances, where the strong correlation plays a bigger role.
The 4c-TCCSD method thus offers benefits compared to single-reference CC methods, however, for future applications the availability of relativistic
4c-CASSCF and/or DMRG-SCF without symmetry restrictions will be important.

\section{APPENDIX}

\subsection{Choice of the DFT functional employed to obtain the molecular spinor basis}

Since the DHF-SCF suffers from a bifurcation of the potential energy curve
and CASSCF was not technically possible for the non-symmetric molecule,
we decided to employ DFT Kohn-Sham spinors as a MO basis for the post-HF calculations.
In order to select a functional, we performed a small scalar quasirelativistic ECP/DFT study of six chosen systems with uranium bonded to terminal nitrogen, where experimental value for $\omega_e$ could be found (Tab.~\ref{DFTstudy}).
We employed three common XC functionals (PBE, BLYP and B3LYP) and def-TZVPP basis set as implemented in the   Turbomole program \cite{TURBOMOLEx},\cite{TURBOMOLEx2}.

\begin{table}[h]
\caption{\label{DFTstudy} DFT functional selection for set of molecules containing N$\equiv$U bond, N$\equiv$U bond stretching frequency $\omega_e$ in $\rm{cm}^{-1}$}
\begin{tabular}{|l|l|r|r|r|r|}
\hline
system  &   exper. & ref & B3LYP &  BLYP &  PBE   \\
\hline
UN$^{+}$\footnotemark[1]{}     &   1083 & \cite{VanGundy_phd}  & 1094  &  1025 &   1058 \\
UN      &   1010  & \cite{KING20142} & 1038  &  977  &   1010  \\
NUO$^{+}$    &   1017 & \cite{KING20142}  & 1182  &  1097 &   1031  \\
NUN\footnotemark[2]{}     &   1051 & \cite{KING20142}  & 1116  &  1054 &   1083  \\
NUF$_3$    &   938 & \cite{KING20142},\cite{Atkinson2018}   & 1042  &  934  &   973  \\
NUNH    &   967 & \cite{KING20142}  & 1038  &  979  &   1013  \\
MAE\footnotemark[4]{}     &       &   &  74   &   32  &    26  \\
MQE\footnotemark[5]{}     &       &   &   90  &    43 &     30 \\
NUHFI   &   n.a.\footnotemark[3]{} &   & 1073  &  974  &   1011 \\
\hline
\end{tabular}
\footnotetext[1]{For this molecular ion, also experimental anharmonicity $\omega_e\rm{x}_e$ = 5.3 $\rm{cm}^{-1}$ has been determined \cite{VanGundy_phd}}
\footnotetext[2]{Antisymmetric stretch ($\sigma_u$), visible in IR spectra}
\footnotetext[3]{To our best knowledge, NUHFI molecule has not been synthesized yet, there are thus no experimental data}
\footnotetext[4]{Mean absolute error}
\footnotetext[5]{Square root of mean of squares of errors}
\end{table}

From the comparison of the computed vibrational frequencies with the experimental ones can be concluded, that best performing functional is PBE.
We have thus selected it as the source of molecular spinors for the post-HF calculations.

\def\drop{drop}
\def\nodrop{nodrop}
\let\reducedmass=\drop

\ifx\reducedmass\nodrop
\subsection{Reduced mass discussion}

In general, to determine the (normal mode specific) reduced mass in a polyatomic molecule, one needs to compute the Hessian in mass-weighted coordinates and diagonalize it. 
Since we confined our study to the single $N-U$ bond stretch (i.e. keeping constant positions of the remaining atoms with respect to Uranium) which only approximates the exact normal vibrational mode, we had to take some
approximation of the corresponding reduced mass. We considered the N-U bond isolated, 
taking $1/\mu =  1/m_{{}^{14}N}+1/m_{U}$.
If we would take the inverse reduced mass of N$\equiv$U stretching mode to be only 1/$m_{{}^{14}N}$ rather than the sum, it would reduce the vibrational frequencies by about 10 cm$^{-1}$, for example the (22,31)TCCSD result for NUF$_{3}$ would become 950 cm$^{-1}$, which is somewhat closer to the experimental value.
While anharmonicities have been computed for the above chosen effective inverse reduced mass $1/\mu =  1/m_{{}^{14}N}+1/m_{U}$ (consistent with computational experiments confirming for ECP/DFT calculation on several uranium nitride molecules that such a choice yields harmonic frequencies with sub cm$^{-1}$ accuracy), frequencies have been evaluated via modifying Hessian from ECP/B3LYP/def-TZVPP approximate calculation in Turbomole according to different second derivative of potential energy cut computed via any method listed in the first column of Spectroscopic parameters Tables (I, II and III) and its rediagonalization. This brings average absolute difference with respect to effective inverse reduced mass method of 3.5 cm$^{-1}$ for NUHFI and 0.45 cm$^{-1}$ for NUF$_{3}$. For NUHFI, methods 4c-CCSD(T), 4c-(20,28)TCCSD and ECP/DFT:r2SCAN
had effective inverse reduced mass method producing outlying difference value with respect to Hessian rediagonalization method of -13 cm$^{-1}$, -14 cm$^{-1}$ and +25 cm$^{-1}$. After removing them from the statistics, the average error is only 1.1 cm$^{-1}$, mostly Hessian rediagonalization giving sligthly higher frequency. 
There is a way to get all modes and anharmonicity simulateously, e.g. Gaussian has module which use simple formulae how from 3 and 4-index tensors of third and fourth derivatives evaluate sec. order. perturbational estimates for matrix of anharmonicities. We could feed it r(2), r(3) and r(4) and either use all values derivatives from Gaussian run or assume all third and fourth derivatives zero, aside for the N-U stretch... but, I would leave it only for the case reviewers might complain...
\luuk{If we extract the Hessian from a calculation with an approximate method, we can potentially modify just the diagonal relating to the motion of the N-atom along the U-N bond. Should be simple if you align that bond along a Cartesian axis. Rediagonalization would then give you a more precise value as then also the slight relative motion of the other atoms is taken into account. I do have a Python script that can do this based on a Hessian taken from ADF, would just need the second derivative in Hartrees/bohr2. Neater is of course to take the Hessian from your Turbomole DFT calculation, but am not sure how easily this is exported ?} Turbomole writes down the Hessian denoted as "projected", I can supplement that data. I will also list the second derivative values for each calculation - along the N-U stretch computed from low oder polynomial fit near minimum
\fi

\begin{acknowledgments}
	The work of the Czech team has been supported by the \textit{Czech Science Foundation} Grant \mbox{No. 18-24563S} and subsequently 
 by the Advanced Multiscale Materials for Key Enabling Technologies project, supported by the Ministry of Education, Youth, and Sports of the Czech Republic. Project No. CZ.02.01.01/00/22\_008/0004558, Co-funded by the European Union.
\mbox{{\"O}. Legeza} has been supported by the Hungarian National Research, Development and Innovation Office (NKFIH) through Grant \mbox{No. K134983} and TKP2021-NVA-04, by  
  the Quantum Information National Laboratory of Hungary, by the Hans Fischer Senior Fellowship programme funded by the Technical University of Munich -- Institute for Advanced Study and by the Center for Scalable and Predictive methods for Excitation and Correlated phenomena (SPEC), funded as part of the Computational Chemical Sciences Program by the U.S.~Department of Energy (DOE), Office of Science, Office of Basic Energy Sciences, Division of Chemical Sciences, Geosciences, and Biosciences at Pacific Northwest National Laboratory
  Mutual visits with the Hungarian group have been partly
	supported by the Hungarian-Czech Joint Research Project MTA/19/04.
	\mbox{J. Brandejs} acknowledges the support of Charles University START programme student grant (No.~CZ.02.2.69/0.0/0.0/19\_073/0016935).
	Computational time for this work was supported by the Ministry of Education, Youth and Sports of the Czech Republic through the e-INFRA CZ (ID:90254). 

\end{acknowledgments}

\bibliographystyle{apsrev4-1}
\bibliography{merged}

\begin{thebibliography}{61}%
\makeatletter
\providecommand \@ifxundefined [1]{%
 \@ifx{#1\undefined}
}%
\providecommand \@ifnum [1]{%
 \ifnum #1\expandafter \@firstoftwo
 \else \expandafter \@secondoftwo
 \fi
}%
\providecommand \@ifx [1]{%
 \ifx #1\expandafter \@firstoftwo
 \else \expandafter \@secondoftwo
 \fi
}%
\providecommand \natexlab [1]{#1}%
\providecommand \enquote  [1]{``#1''}%
\providecommand \bibnamefont  [1]{#1}%
\providecommand \bibfnamefont [1]{#1}%
\providecommand \citenamefont [1]{#1}%
\providecommand \href@noop [0]{\@secondoftwo}%
\providecommand \href [0]{\begingroup \@sanitize@url \@href}%
\providecommand \@href[1]{\@@startlink{#1}\@@href}%
\providecommand \@@href[1]{\endgroup#1\@@endlink}%
\providecommand \@sanitize@url [0]{\catcode `\\12\catcode `\$12\catcode
  `\&12\catcode `\#12\catcode `\^12\catcode `\_12\catcode `\%12\relax}%
\providecommand \@@startlink[1]{}%
\providecommand \@@endlink[0]{}%
\providecommand \url  [0]{\begingroup\@sanitize@url \@url }%
\providecommand \@url [1]{\endgroup\@href {#1}{\urlprefix }}%
\providecommand \urlprefix  [0]{URL }%
\providecommand \Eprint [0]{\href }%
\providecommand \doibase [0]{http://dx.doi.org/}%
\providecommand \selectlanguage [0]{\@gobble}%
\providecommand \bibinfo  [0]{\@secondoftwo}%
\providecommand \bibfield  [0]{\@secondoftwo}%
\providecommand \translation [1]{[#1]}%
\providecommand \BibitemOpen [0]{}%
\providecommand \bibitemStop [0]{}%
\providecommand \bibitemNoStop [0]{.\EOS\space}%
\providecommand \EOS [0]{\spacefactor3000\relax}%
\providecommand \BibitemShut  [1]{\csname bibitem#1\endcsname}%
\let\auto@bib@innerbib\@empty
\bibitem [{\citenamefont {White}(1992)}]{White1992}%
  \BibitemOpen
  \bibfield  {author} {\bibinfo {author} {\bibfnamefont {S.~R.}\ \bibnamefont
  {White}},\ }\href {\doibase 10.1103/physrevlett.69.2863} {\bibfield
  {journal} {\bibinfo  {journal} {Physical Review Letters}\ }\textbf {\bibinfo
  {volume} {69}},\ \bibinfo {pages} {2863} (\bibinfo {year}
  {1992})}\BibitemShut {NoStop}%
\bibitem [{\citenamefont {White}\ and\ \citenamefont
  {Martin}(1999)}]{White1999}%
  \BibitemOpen
  \bibfield  {author} {\bibinfo {author} {\bibfnamefont {S.~R.}\ \bibnamefont
  {White}}\ and\ \bibinfo {author} {\bibfnamefont {R.~L.}\ \bibnamefont
  {Martin}},\ }\href {\doibase 10.1063/1.478295} {\bibfield  {journal}
  {\bibinfo  {journal} {The Journal of Chemical Physics}\ }\textbf {\bibinfo
  {volume} {110}},\ \bibinfo {pages} {4127} (\bibinfo {year}
  {1999})}\BibitemShut {NoStop}%
\bibitem [{\citenamefont {Chan}\ and\ \citenamefont
  {Head-Gordon}(2002)}]{Chan2002}%
  \BibitemOpen
  \bibfield  {author} {\bibinfo {author} {\bibfnamefont {G.~K.-L.}\
  \bibnamefont {Chan}}\ and\ \bibinfo {author} {\bibfnamefont {M.}~\bibnamefont
  {Head-Gordon}},\ }\href {\doibase 10.1063/1.1449459} {\bibfield  {journal}
  {\bibinfo  {journal} {The Journal of Chemical Physics}\ }\textbf {\bibinfo
  {volume} {116}},\ \bibinfo {pages} {4462} (\bibinfo {year}
  {2002})}\BibitemShut {NoStop}%
\bibitem [{\citenamefont {Legeza}\ \emph {et~al.}(2003)\citenamefont {Legeza},
  \citenamefont {R\"{o}der},\ and\ \citenamefont {Hess}}]{Legeza2003a}%
  \BibitemOpen
  \bibfield  {author} {\bibinfo {author} {\bibfnamefont {{\"O}.}~\bibnamefont
  {Legeza}}, \bibinfo {author} {\bibfnamefont {J.}~\bibnamefont {R\"{o}der}}, \
  and\ \bibinfo {author} {\bibfnamefont {B.~A.}\ \bibnamefont {Hess}},\
  }\href@noop {} {\bibfield  {journal} {\bibinfo  {journal} {Physical Review
  B}\ }\textbf {\bibinfo {volume} {67}},\ \bibinfo {pages} {125114} (\bibinfo
  {year} {2003})}\BibitemShut {NoStop}%
\bibitem [{\citenamefont {Schollw\"{o}ck}(2011)}]{Schollwoeck2011}%
  \BibitemOpen
  \bibfield  {author} {\bibinfo {author} {\bibfnamefont {U.}~\bibnamefont
  {Schollw\"{o}ck}},\ }\href {\doibase 10.1016/j.aop.2010.09.012} {\bibfield
  {journal} {\bibinfo  {journal} {Annals of Physics}\ }\textbf {\bibinfo
  {volume} {326}},\ \bibinfo {pages} {96} (\bibinfo {year} {2011})}\BibitemShut
  {NoStop}%
\bibitem [{\citenamefont {Olivares-Amaya}\ \emph {et~al.}(2015)\citenamefont
  {Olivares-Amaya}, \citenamefont {Hu}, \citenamefont {Nakatani}, \citenamefont
  {Sharma}, \citenamefont {Yang},\ and\ \citenamefont
  {Chan}}]{OlivaresAmaya2015}%
  \BibitemOpen
  \bibfield  {author} {\bibinfo {author} {\bibfnamefont {R.}~\bibnamefont
  {Olivares-Amaya}}, \bibinfo {author} {\bibfnamefont {W.}~\bibnamefont {Hu}},
  \bibinfo {author} {\bibfnamefont {N.}~\bibnamefont {Nakatani}}, \bibinfo
  {author} {\bibfnamefont {S.}~\bibnamefont {Sharma}}, \bibinfo {author}
  {\bibfnamefont {J.}~\bibnamefont {Yang}}, \ and\ \bibinfo {author}
  {\bibfnamefont {G.~K.-L.}\ \bibnamefont {Chan}},\ }\href {\doibase
  10.1063/1.4905329} {\bibfield  {journal} {\bibinfo  {journal} {The Journal of
  Chemical Physics}\ }\textbf {\bibinfo {volume} {142}},\ \bibinfo {pages}
  {034102} (\bibinfo {year} {2015})}\BibitemShut {NoStop}%
\bibitem [{\citenamefont {Kurashige}\ and\ \citenamefont
  {Yanai}(2011)}]{Kurashige2011}%
  \BibitemOpen
  \bibfield  {author} {\bibinfo {author} {\bibfnamefont {Y.}~\bibnamefont
  {Kurashige}}\ and\ \bibinfo {author} {\bibfnamefont {T.}~\bibnamefont
  {Yanai}},\ }\href {\doibase 10.1063/1.3629454} {\bibfield  {journal}
  {\bibinfo  {journal} {The Journal of Chemical Physics}\ }\textbf {\bibinfo
  {volume} {135}},\ \bibinfo {pages} {094104} (\bibinfo {year}
  {2011})}\BibitemShut {NoStop}%
\bibitem [{\citenamefont {Saitow}\ \emph {et~al.}(2013)\citenamefont {Saitow},
  \citenamefont {Kurashige},\ and\ \citenamefont {Yanai}}]{Saitow2013}%
  \BibitemOpen
  \bibfield  {author} {\bibinfo {author} {\bibfnamefont {M.}~\bibnamefont
  {Saitow}}, \bibinfo {author} {\bibfnamefont {Y.}~\bibnamefont {Kurashige}}, \
  and\ \bibinfo {author} {\bibfnamefont {T.}~\bibnamefont {Yanai}},\ }\href
  {\doibase 10.1063/1.4816627} {\bibfield  {journal} {\bibinfo  {journal} {The
  Journal of Chemical Physics}\ }\textbf {\bibinfo {volume} {139}},\ \bibinfo
  {pages} {044118} (\bibinfo {year} {2013})}\BibitemShut {NoStop}%
\bibitem [{\citenamefont {Wouters}\ \emph {et~al.}(2013)\citenamefont
  {Wouters}, \citenamefont {Nakatani}, \citenamefont {Neck},\ and\
  \citenamefont {Chan}}]{Wouters2013}%
  \BibitemOpen
  \bibfield  {author} {\bibinfo {author} {\bibfnamefont {S.}~\bibnamefont
  {Wouters}}, \bibinfo {author} {\bibfnamefont {N.}~\bibnamefont {Nakatani}},
  \bibinfo {author} {\bibfnamefont {D.~V.}\ \bibnamefont {Neck}}, \ and\
  \bibinfo {author} {\bibfnamefont {G.~K.-L.}\ \bibnamefont {Chan}},\
  }\href@noop {} {\bibfield  {journal} {\bibinfo  {journal} {Physical Review
  B}\ }\textbf {\bibinfo {volume} {88}},\ \bibinfo {pages} {075122} (\bibinfo
  {year} {2013})}\BibitemShut {NoStop}%
\bibitem [{\citenamefont {Yanai}\ and\ \citenamefont {Chan}(2006)}]{Yanai2006}%
  \BibitemOpen
  \bibfield  {author} {\bibinfo {author} {\bibfnamefont {T.}~\bibnamefont
  {Yanai}}\ and\ \bibinfo {author} {\bibfnamefont {G.~K.-L.}\ \bibnamefont
  {Chan}},\ }\href {\doibase 10.1063/1.2196410} {\bibfield  {journal} {\bibinfo
   {journal} {The Journal of Chemical Physics}\ }\textbf {\bibinfo {volume}
  {124}},\ \bibinfo {pages} {194106} (\bibinfo {year} {2006})}\BibitemShut
  {NoStop}%
\bibitem [{\citenamefont {Ren}\ \emph {et~al.}(2016)\citenamefont {Ren},
  \citenamefont {Yi},\ and\ \citenamefont {Shuai}}]{Ren2016}%
  \BibitemOpen
  \bibfield  {author} {\bibinfo {author} {\bibfnamefont {J.}~\bibnamefont
  {Ren}}, \bibinfo {author} {\bibfnamefont {Y.}~\bibnamefont {Yi}}, \ and\
  \bibinfo {author} {\bibfnamefont {Z.}~\bibnamefont {Shuai}},\ }\href
  {\doibase 10.1021/acs.jctc.6b00696} {\bibfield  {journal} {\bibinfo
  {journal} {Journal of Chemical Theory and Computation}\ }\textbf {\bibinfo
  {volume} {12}},\ \bibinfo {pages} {4871} (\bibinfo {year}
  {2016})}\BibitemShut {NoStop}%
\bibitem [{\citenamefont {Beran}\ \emph {et~al.}(2021)\citenamefont {Beran},
  \citenamefont {Matou\v{s}ek}, \citenamefont {Hapka}, \citenamefont {Pernal},\
  and\ \citenamefont {Veis}}]{Veis-Pernal2021}%
  \BibitemOpen
  \bibfield  {author} {\bibinfo {author} {\bibfnamefont {P.}~\bibnamefont
  {Beran}}, \bibinfo {author} {\bibfnamefont {M.}~\bibnamefont {Matou\v{s}ek}},
  \bibinfo {author} {\bibfnamefont {M.}~\bibnamefont {Hapka}}, \bibinfo
  {author} {\bibfnamefont {K.}~\bibnamefont {Pernal}}, \ and\ \bibinfo {author}
  {\bibfnamefont {L.}~\bibnamefont {Veis}},\ }\href@noop {} {\bibfield
  {journal} {\bibinfo  {journal} {arXiv}\ ,\ \bibinfo {pages} {2108.12803}}
  (\bibinfo {year} {2021})}\BibitemShut {NoStop}%
\bibitem [{\citenamefont {Barcza}\ \emph {et~al.}(2022)\citenamefont {Barcza},
  \citenamefont {Werner}, \citenamefont {Zar{\'{a}}nd}, \citenamefont
  {Pershin}, \citenamefont {Benedek}, \citenamefont {Örs Legeza},\ and\
  \citenamefont {Szilv{\'{a}}si}}]{Barcza-2022b}%
  \BibitemOpen
  \bibfield  {author} {\bibinfo {author} {\bibfnamefont {G.}~\bibnamefont
  {Barcza}}, \bibinfo {author} {\bibfnamefont {M.~A.}\ \bibnamefont {Werner}},
  \bibinfo {author} {\bibfnamefont {G.}~\bibnamefont {Zar{\'{a}}nd}}, \bibinfo
  {author} {\bibfnamefont {A.}~\bibnamefont {Pershin}}, \bibinfo {author}
  {\bibfnamefont {Z.}~\bibnamefont {Benedek}}, \bibinfo {author} {\bibnamefont
  {Örs Legeza}}, \ and\ \bibinfo {author} {\bibfnamefont {T.}~\bibnamefont
  {Szilv{\'{a}}si}},\ }\href {\doibase 10.1021/acs.jpca.2c05952} {\bibfield
  {journal} {\bibinfo  {journal} {The Journal of Physical Chemistry A}\
  }\textbf {\bibinfo {volume} {126}},\ \bibinfo {pages} {9709} (\bibinfo {year}
  {2022})}\BibitemShut {NoStop}%
\bibitem [{\citenamefont {Friesecke}\ \emph {et~al.}(2023)\citenamefont
  {Friesecke}, \citenamefont {Barcza},\ and\ \citenamefont
  {Legeza}}]{Friesecke-2023}%
  \BibitemOpen
  \bibfield  {author} {\bibinfo {author} {\bibfnamefont {G.}~\bibnamefont
  {Friesecke}}, \bibinfo {author} {\bibfnamefont {G.}~\bibnamefont {Barcza}}, \
  and\ \bibinfo {author} {\bibfnamefont {O.}~\bibnamefont {Legeza}},\ }\href
  {https://doi.org/10.1021/acs.jctc.3c01001} {\bibfield  {journal} {\bibinfo
  {journal} {Journal of Chemical Theory and Computation}\ }\textbf {\bibinfo
  {volume} {20}},\ \bibinfo {pages} {87} (\bibinfo {year} {2023})}\BibitemShut
  {NoStop}%
\bibitem [{\citenamefont {Veis}\ \emph {et~al.}(2016)\citenamefont {Veis},
  \citenamefont {Antal{\'{\i}}k}, \citenamefont {Brabec}, \citenamefont
  {Neese}, \citenamefont {Legeza},\ and\ \citenamefont {Pittner}}]{Veis2016}%
  \BibitemOpen
  \bibfield  {author} {\bibinfo {author} {\bibfnamefont {L.}~\bibnamefont
  {Veis}}, \bibinfo {author} {\bibfnamefont {A.}~\bibnamefont
  {Antal{\'{\i}}k}}, \bibinfo {author} {\bibfnamefont {J.}~\bibnamefont
  {Brabec}}, \bibinfo {author} {\bibfnamefont {F.}~\bibnamefont {Neese}},
  \bibinfo {author} {\bibfnamefont {{\"O}.}~\bibnamefont {Legeza}}, \ and\
  \bibinfo {author} {\bibfnamefont {J.}~\bibnamefont {Pittner}},\ }\href
  {\doibase 10.1021/acs.jpclett.6b01908} {\bibfield  {journal} {\bibinfo
  {journal} {The Journal of Physical Chemistry Letters}\ }\textbf {\bibinfo
  {volume} {7}},\ \bibinfo {pages} {4072} (\bibinfo {year} {2016})}\BibitemShut
  {NoStop}%
\bibitem [{\citenamefont {Kinoshita}\ \emph {et~al.}(2005)\citenamefont
  {Kinoshita}, \citenamefont {Hino},\ and\ \citenamefont
  {Bartlett}}]{Kinoshita2005}%
  \BibitemOpen
  \bibfield  {author} {\bibinfo {author} {\bibfnamefont {T.}~\bibnamefont
  {Kinoshita}}, \bibinfo {author} {\bibfnamefont {O.}~\bibnamefont {Hino}}, \
  and\ \bibinfo {author} {\bibfnamefont {R.~J.}\ \bibnamefont {Bartlett}},\
  }\href {\doibase 10.1063/1.2000251} {\bibfield  {journal} {\bibinfo
  {journal} {The Journal of Chemical Physics}\ }\textbf {\bibinfo {volume}
  {123}},\ \bibinfo {pages} {074106} (\bibinfo {year} {2005})}\BibitemShut
  {NoStop}%
\bibitem [{\citenamefont {Hino}\ \emph {et~al.}(2006)\citenamefont {Hino},
  \citenamefont {Kinoshita}, \citenamefont {Chan},\ and\ \citenamefont
  {Bartlett}}]{Hino2006}%
  \BibitemOpen
  \bibfield  {author} {\bibinfo {author} {\bibfnamefont {O.}~\bibnamefont
  {Hino}}, \bibinfo {author} {\bibfnamefont {T.}~\bibnamefont {Kinoshita}},
  \bibinfo {author} {\bibfnamefont {G.~K.-L.}\ \bibnamefont {Chan}}, \ and\
  \bibinfo {author} {\bibfnamefont {R.~J.}\ \bibnamefont {Bartlett}},\ }\href
  {\doibase 10.1063/1.2180775} {\bibfield  {journal} {\bibinfo  {journal} {The
  Journal of Chemical Physics}\ }\textbf {\bibinfo {volume} {124}},\ \bibinfo
  {pages} {114311} (\bibinfo {year} {2006})}\BibitemShut {NoStop}%
\bibitem [{\citenamefont {Veis}\ \emph {et~al.}(2017)\citenamefont {Veis},
  \citenamefont {Antal{\'{\i}}k}, \citenamefont {Brabec}, \citenamefont
  {Neese}, \citenamefont {Legeza},\ and\ \citenamefont
  {Pittner}}]{Veis2016err}%
  \BibitemOpen
  \bibfield  {author} {\bibinfo {author} {\bibfnamefont {L.}~\bibnamefont
  {Veis}}, \bibinfo {author} {\bibfnamefont {A.}~\bibnamefont
  {Antal{\'{\i}}k}}, \bibinfo {author} {\bibfnamefont {J.}~\bibnamefont
  {Brabec}}, \bibinfo {author} {\bibfnamefont {F.}~\bibnamefont {Neese}},
  \bibinfo {author} {\bibfnamefont {{\"O}.}~\bibnamefont {Legeza}}, \ and\
  \bibinfo {author} {\bibfnamefont {J.}~\bibnamefont {Pittner}},\ }\href@noop
  {} {\bibfield  {journal} {\bibinfo  {journal} {The Journal of Physical
  Chemistry Letters}\ }\textbf {\bibinfo {volume} {8}},\ \bibinfo {pages} {291}
  (\bibinfo {year} {2017})}\BibitemShut {NoStop}%
\bibitem [{\citenamefont {Faulstich}\ \emph
  {et~al.}(2019{\natexlab{a}})\citenamefont {Faulstich}, \citenamefont
  {Laestadius}, \citenamefont {\"{O}rs Legeza}, \citenamefont {Schneider},\
  and\ \citenamefont {Kvaal}}]{Faulstich2019a}%
  \BibitemOpen
  \bibfield  {author} {\bibinfo {author} {\bibfnamefont {F.~M.}\ \bibnamefont
  {Faulstich}}, \bibinfo {author} {\bibfnamefont {A.}~\bibnamefont
  {Laestadius}}, \bibinfo {author} {\bibnamefont {\"{O}rs Legeza}}, \bibinfo
  {author} {\bibfnamefont {R.}~\bibnamefont {Schneider}}, \ and\ \bibinfo
  {author} {\bibfnamefont {S.}~\bibnamefont {Kvaal}},\ }\href {\doibase
  10.1137/18m1171436} {\bibfield  {journal} {\bibinfo  {journal} {{SIAM}
  Journal on Numerical Analysis}\ }\textbf {\bibinfo {volume} {57}},\ \bibinfo
  {pages} {2579} (\bibinfo {year} {2019}{\natexlab{a}})}\BibitemShut {NoStop}%
\bibitem [{\citenamefont {Faulstich}\ \emph
  {et~al.}(2019{\natexlab{b}})\citenamefont {Faulstich}, \citenamefont
  {M{\'{a}}t{\'{e}}}, \citenamefont {Laestadius}, \citenamefont {Csirik},
  \citenamefont {Veis}, \citenamefont {Antalik}, \citenamefont {Brabec},
  \citenamefont {Schneider}, \citenamefont {Pittner}, \citenamefont {Kvaal},\
  and\ \citenamefont {Legeza}}]{Faulstich2019b}%
  \BibitemOpen
  \bibfield  {author} {\bibinfo {author} {\bibfnamefont {F.~M.}\ \bibnamefont
  {Faulstich}}, \bibinfo {author} {\bibfnamefont {M.}~\bibnamefont
  {M{\'{a}}t{\'{e}}}}, \bibinfo {author} {\bibfnamefont {A.}~\bibnamefont
  {Laestadius}}, \bibinfo {author} {\bibfnamefont {M.~A.}\ \bibnamefont
  {Csirik}}, \bibinfo {author} {\bibfnamefont {L.}~\bibnamefont {Veis}},
  \bibinfo {author} {\bibfnamefont {A.}~\bibnamefont {Antalik}}, \bibinfo
  {author} {\bibfnamefont {J.}~\bibnamefont {Brabec}}, \bibinfo {author}
  {\bibfnamefont {R.}~\bibnamefont {Schneider}}, \bibinfo {author}
  {\bibfnamefont {J.}~\bibnamefont {Pittner}}, \bibinfo {author} {\bibfnamefont
  {S.}~\bibnamefont {Kvaal}}, \ and\ \bibinfo {author} {\bibfnamefont
  {{\"O}.}~\bibnamefont {Legeza}},\ }\href {\doibase 10.1021/acs.jctc.8b00960}
  {\bibfield  {journal} {\bibinfo  {journal} {Journal of Chemical Theory and
  Computation}\ }\textbf {\bibinfo {volume} {15}},\ \bibinfo {pages} {2206}
  (\bibinfo {year} {2019}{\natexlab{b}})}\BibitemShut {NoStop}%
\bibitem [{\citenamefont {Leszczyk}\ \emph {et~al.}(2022)\citenamefont
  {Leszczyk}, \citenamefont {Máté}, \citenamefont {Legeza},\ and\
  \citenamefont {Boguslawski}}]{Leszczyk-2022}%
  \BibitemOpen
  \bibfield  {author} {\bibinfo {author} {\bibfnamefont {A.}~\bibnamefont
  {Leszczyk}}, \bibinfo {author} {\bibfnamefont {M.}~\bibnamefont {Máté}},
  \bibinfo {author} {\bibfnamefont {{\"O}.}~\bibnamefont {Legeza}}, \ and\
  \bibinfo {author} {\bibfnamefont {K.}~\bibnamefont {Boguslawski}},\
  }\href@noop {} {\bibfield  {journal} {\bibinfo  {journal} {J. Chem. Theory
  Comp.}\ }\textbf {\bibinfo {volume} {18}},\ \bibinfo {pages} {96} (\bibinfo
  {year} {2022})}\BibitemShut {NoStop}%
\bibitem [{\citenamefont {Neese}(2012)}]{Neese2012}%
  \BibitemOpen
  \bibfield  {author} {\bibinfo {author} {\bibfnamefont {F.}~\bibnamefont
  {Neese}},\ }\href {\doibase https://doi.org/10.1002/wcms.81} {\bibfield
  {journal} {\bibinfo  {journal} {WIREs Computational Molecular Science}\
  }\textbf {\bibinfo {volume} {2}},\ \bibinfo {pages} {73} (\bibinfo {year}
  {2012})}\BibitemShut {NoStop}%
\bibitem [{\citenamefont {Antalik}\ \emph {et~al.}(2019)\citenamefont
  {Antalik}, \citenamefont {Veis}, \citenamefont {Brabec}, \citenamefont
  {Demel}, \citenamefont {Legeza},\ and\ \citenamefont {Pittner}}]{lpno-tcc}%
  \BibitemOpen
  \bibfield  {author} {\bibinfo {author} {\bibfnamefont {A.}~\bibnamefont
  {Antalik}}, \bibinfo {author} {\bibfnamefont {L.}~\bibnamefont {Veis}},
  \bibinfo {author} {\bibfnamefont {J.}~\bibnamefont {Brabec}}, \bibinfo
  {author} {\bibfnamefont {O.}~\bibnamefont {Demel}}, \bibinfo {author}
  {\bibfnamefont {O.}~\bibnamefont {Legeza}}, \ and\ \bibinfo {author}
  {\bibfnamefont {J.}~\bibnamefont {Pittner}},\ }\href@noop {} {\bibfield
  {journal} {\bibinfo  {journal} {J. Chem. Phys.}\ }\textbf {\bibinfo {volume}
  {151}},\ \bibinfo {pages} {084112} (\bibinfo {year} {2019})}\BibitemShut
  {NoStop}%
\bibitem [{\citenamefont {Lang}\ \emph {et~al.}(2020)\citenamefont {Lang},
  \citenamefont {Antalik}, \citenamefont {Veis}, \citenamefont {Brabec},
  \citenamefont {Legeza},\ and\ \citenamefont {Pittner}}]{dlpno-tcc}%
  \BibitemOpen
  \bibfield  {author} {\bibinfo {author} {\bibfnamefont {J.}~\bibnamefont
  {Lang}}, \bibinfo {author} {\bibfnamefont {A.}~\bibnamefont {Antalik}},
  \bibinfo {author} {\bibfnamefont {L.}~\bibnamefont {Veis}}, \bibinfo {author}
  {\bibfnamefont {J.}~\bibnamefont {Brabec}}, \bibinfo {author} {\bibfnamefont
  {O.}~\bibnamefont {Legeza}}, \ and\ \bibinfo {author} {\bibfnamefont
  {J.}~\bibnamefont {Pittner}},\ }\href@noop {} {\bibfield  {journal} {\bibinfo
   {journal} {J. Chem. Theor. Comp.}\ }\textbf {\bibinfo {volume} {16}},\
  \bibinfo {pages} {3028} (\bibinfo {year} {2020})}\BibitemShut {NoStop}%
\bibitem [{\citenamefont {Antalik}\ \emph {et~al.}(2020)\citenamefont
  {Antalik}, \citenamefont {Nachtigallova}, \citenamefont {Lo}, \citenamefont
  {Matousek}, \citenamefont {LAng}, \citenamefont {Legeza}, \citenamefont
  {Pittner}, \citenamefont {Hobza},\ and\ \citenamefont
  {Veis}}]{dlpno-tcc-app}%
  \BibitemOpen
  \bibfield  {author} {\bibinfo {author} {\bibfnamefont {A.}~\bibnamefont
  {Antalik}}, \bibinfo {author} {\bibfnamefont {D.}~\bibnamefont
  {Nachtigallova}}, \bibinfo {author} {\bibfnamefont {R.}~\bibnamefont {Lo}},
  \bibinfo {author} {\bibfnamefont {M.}~\bibnamefont {Matousek}}, \bibinfo
  {author} {\bibfnamefont {J.}~\bibnamefont {LAng}}, \bibinfo {author}
  {\bibfnamefont {O.}~\bibnamefont {Legeza}}, \bibinfo {author} {\bibfnamefont
  {J.}~\bibnamefont {Pittner}}, \bibinfo {author} {\bibfnamefont
  {P.}~\bibnamefont {Hobza}}, \ and\ \bibinfo {author} {\bibfnamefont
  {L.}~\bibnamefont {Veis}},\ }\href@noop {} {\bibfield  {journal} {\bibinfo
  {journal} {Phys. Chem. Chem. Phys.}\ }\textbf {\bibinfo {volume} {22}},\
  \bibinfo {pages} {17033} (\bibinfo {year} {2020})}\BibitemShut {NoStop}%
\bibitem [{\citenamefont {Brandejs}\ \emph {et~al.}(2020)\citenamefont
  {Brandejs}, \citenamefont {Vi\v{s}\v{n}\'{a}k}, \citenamefont {Veis},
  \citenamefont {Mat\'{e}}, \citenamefont {Legeza},\ and\ \citenamefont
  {Pittner}}]{reltcc2020}%
  \BibitemOpen
  \bibfield  {author} {\bibinfo {author} {\bibfnamefont {J.}~\bibnamefont
  {Brandejs}}, \bibinfo {author} {\bibfnamefont {J.}~\bibnamefont
  {Vi\v{s}\v{n}\'{a}k}}, \bibinfo {author} {\bibfnamefont {L.}~\bibnamefont
  {Veis}}, \bibinfo {author} {\bibfnamefont {M.}~\bibnamefont {Mat\'{e}}},
  \bibinfo {author} {\bibfnamefont {O.}~\bibnamefont {Legeza}}, \ and\ \bibinfo
  {author} {\bibfnamefont {J.}~\bibnamefont {Pittner}},\ }\href@noop {}
  {\bibfield  {journal} {\bibinfo  {journal} {The Journal of Chemical Physics}\
  }\textbf {\bibinfo {volume} {152}},\ \bibinfo {pages} {174107} (\bibinfo
  {year} {2020})}\BibitemShut {NoStop}%
\bibitem [{\citenamefont {Wormit}\ \emph {et~al.}(2014)\citenamefont {Wormit},
  \citenamefont {Olejniczak}, \citenamefont {Deppenmeier}, \citenamefont
  {Borschevsky}, \citenamefont {Saue},\ and\ \citenamefont
  {PeterSchwerdtfeger}}]{Wormit2014}%
  \BibitemOpen
  \bibfield  {author} {\bibinfo {author} {\bibfnamefont {M.}~\bibnamefont
  {Wormit}}, \bibinfo {author} {\bibfnamefont {M.}~\bibnamefont {Olejniczak}},
  \bibinfo {author} {\bibfnamefont {A.-L.}\ \bibnamefont {Deppenmeier}},
  \bibinfo {author} {\bibfnamefont {A.}~\bibnamefont {Borschevsky}}, \bibinfo
  {author} {\bibfnamefont {T.}~\bibnamefont {Saue}}, \ and\ \bibinfo {author}
  {\bibnamefont {PeterSchwerdtfeger}},\ }\href@noop {} {\bibfield  {journal}
  {\bibinfo  {journal} {Phys. Chem. Chem. Phys.}\ }\textbf {\bibinfo {volume}
  {16}},\ \bibinfo {pages} {17043} (\bibinfo {year} {2014})}\BibitemShut
  {NoStop}%
\bibitem [{\citenamefont {Jiang}\ and\ \citenamefont
  {Wilson}(2011)}]{Jiang2011}%
  \BibitemOpen
  \bibfield  {author} {\bibinfo {author} {\bibfnamefont {W.}~\bibnamefont
  {Jiang}}\ and\ \bibinfo {author} {\bibfnamefont {A.~K.}\ \bibnamefont
  {Wilson}},\ }\href {\doibase 10.1063/1.3514031} {\bibfield  {journal}
  {\bibinfo  {journal} {The Journal of Chemical Physics}\ }\textbf {\bibinfo
  {volume} {134}},\ \bibinfo {pages} {034101} (\bibinfo {year}
  {2011})}\BibitemShut {NoStop}%
\bibitem [{\citenamefont {Mintz}\ \emph {et~al.}(2009)\citenamefont {Mintz},
  \citenamefont {Williams}, \citenamefont {Howard},\ and\ \citenamefont
  {Wilson}}]{Mintz2009}%
  \BibitemOpen
  \bibfield  {author} {\bibinfo {author} {\bibfnamefont {B.}~\bibnamefont
  {Mintz}}, \bibinfo {author} {\bibfnamefont {T.~G.}\ \bibnamefont {Williams}},
  \bibinfo {author} {\bibfnamefont {L.}~\bibnamefont {Howard}}, \ and\ \bibinfo
  {author} {\bibfnamefont {A.~K.}\ \bibnamefont {Wilson}},\ }\href {\doibase
  10.1063/1.3149387} {\bibfield  {journal} {\bibinfo  {journal} {The Journal of
  Chemical Physics}\ }\textbf {\bibinfo {volume} {130}},\ \bibinfo {pages}
  {234104} (\bibinfo {year} {2009})}\BibitemShut {NoStop}%
\bibitem [{\citenamefont {Chan}\ \emph {et~al.}(2004)\citenamefont {Chan},
  \citenamefont {K\'allay},\ and\ \citenamefont {Gauss}}]{Chan-2004b}%
  \BibitemOpen
  \bibfield  {author} {\bibinfo {author} {\bibfnamefont {G.~K.-L.}\
  \bibnamefont {Chan}}, \bibinfo {author} {\bibfnamefont {M.}~\bibnamefont
  {K\'allay}}, \ and\ \bibinfo {author} {\bibfnamefont {J.}~\bibnamefont
  {Gauss}},\ }\href {\doibase http://dx.doi.org/10.1063/1.1783212} {\bibfield
  {journal} {\bibinfo  {journal} {The Journal of Chemical Physics}\ }\textbf
  {\bibinfo {volume} {121}},\ \bibinfo {pages} {6110} (\bibinfo {year}
  {2004})}\BibitemShut {NoStop}%
\bibitem [{\citenamefont {Máté}\ \emph {et~al.}(2023)\citenamefont {Máté},
  \citenamefont {Petrov}, \citenamefont {Szalay},\ and\ \citenamefont
  {Legeza}}]{Mate-2022}%
  \BibitemOpen
  \bibfield  {author} {\bibinfo {author} {\bibfnamefont {M.}~\bibnamefont
  {Máté}}, \bibinfo {author} {\bibfnamefont {K.}~\bibnamefont {Petrov}},
  \bibinfo {author} {\bibfnamefont {S.}~\bibnamefont {Szalay}}, \ and\ \bibinfo
  {author} {\bibfnamefont {{\"O}.}~\bibnamefont {Legeza}},\ }\href@noop {}
  {\bibfield  {journal} {\bibinfo  {journal} {Journal of Mathematical
  Chemistry}\ }\textbf {\bibinfo {volume} {61}},\ \bibinfo {pages} {362}
  (\bibinfo {year} {2023})}\BibitemShut {NoStop}%
\bibitem [{\citenamefont {Li}\ and\ \citenamefont
  {Paldus}(1998)}]{LiPaldus1998}%
  \BibitemOpen
  \bibfield  {author} {\bibinfo {author} {\bibfnamefont {X.}~\bibnamefont
  {Li}}\ and\ \bibinfo {author} {\bibfnamefont {J.}~\bibnamefont {Paldus}},\
  }\href {\doibase https://doi.org/10.1016/S0009-2614(97)01132-9} {\bibfield
  {journal} {\bibinfo  {journal} {Chemical Physics Letters}\ }\textbf {\bibinfo
  {volume} {286}},\ \bibinfo {pages} {145} (\bibinfo {year}
  {1998})}\BibitemShut {NoStop}%
\bibitem [{\citenamefont {Kramers}(1930)}]{KRAMERS1930}%
  \BibitemOpen
  \bibfield  {author} {\bibinfo {author} {\bibfnamefont {H.~A.}\ \bibnamefont
  {Kramers}},\ }\href@noop {} {\bibfield  {journal} {\bibinfo  {journal} {Proc.
  Acad. Amst}\ } (\bibinfo {year} {1930})}\BibitemShut {NoStop}%
\bibitem [{\citenamefont {Fleig}\ \emph {et~al.}(2001)\citenamefont {Fleig},
  \citenamefont {Olsen},\ and\ \citenamefont {Marian}}]{fleig2001}%
  \BibitemOpen
  \bibfield  {author} {\bibinfo {author} {\bibfnamefont {T.}~\bibnamefont
  {Fleig}}, \bibinfo {author} {\bibfnamefont {J.}~\bibnamefont {Olsen}}, \ and\
  \bibinfo {author} {\bibfnamefont {C.~M.}\ \bibnamefont {Marian}},\ }\href
  {\doibase 10.1063/1.1349076} {\bibfield  {journal} {\bibinfo  {journal} {The
  Journal of Chemical Physics}\ }\textbf {\bibinfo {volume} {114}},\ \bibinfo
  {pages} {4775} (\bibinfo {year} {2001})}\BibitemShut {NoStop}%
\bibitem [{DIR(2018)}]{DIRAC18}%
  \BibitemOpen
  \href@noop {} {} (\bibinfo {year} {2018}),\ \bibinfo {note} {{DIRAC}, a
  relativistic ab initio electronic structure program (2018), T.~Saue et al.
  \url{http://www.diracprogram.org}}\BibitemShut {NoStop}%
\bibitem [{\citenamefont {Saue}\ and\ \citenamefont
  {Jensen}(1999)}]{saue-jensen-1999}%
  \BibitemOpen
  \bibfield  {author} {\bibinfo {author} {\bibfnamefont {T.}~\bibnamefont
  {Saue}}\ and\ \bibinfo {author} {\bibfnamefont {H.~J.~A.}\ \bibnamefont
  {Jensen}},\ }\href {\doibase 10.1063/1.479958} {\bibfield  {journal}
  {\bibinfo  {journal} {The Journal of Chemical Physics}\ }\textbf {\bibinfo
  {volume} {111}},\ \bibinfo {pages} {6211} (\bibinfo {year}
  {1999})}\BibitemShut {NoStop}%
\bibitem [{\citenamefont {Dyall}\ and\ \citenamefont
  {F{\ae}gri~Jr}(2007)}]{dyall-faegri}%
  \BibitemOpen
  \bibfield  {author} {\bibinfo {author} {\bibfnamefont {K.~G.}\ \bibnamefont
  {Dyall}}\ and\ \bibinfo {author} {\bibfnamefont {K.}~\bibnamefont
  {F{\ae}gri~Jr}},\ }\href@noop {} {\emph {\bibinfo {title} {Introduction to
  relativistic quantum chemistry}}}\ (\bibinfo  {publisher} {Oxford University
  Press},\ \bibinfo {year} {2007})\BibitemShut {NoStop}%
\bibitem [{\citenamefont {Visscher}(1996)}]{Visscher1996}%
  \BibitemOpen
  \bibfield  {author} {\bibinfo {author} {\bibfnamefont {L.}~\bibnamefont
  {Visscher}},\ }\href {<Go to ISI>://A1996UK63500004} {\bibfield  {journal}
  {\bibinfo  {journal} {Chemical Physics Letters}\ }\textbf {\bibinfo {volume}
  {253}},\ \bibinfo {pages} {20 } (\bibinfo {year} {1996})}\BibitemShut
  {NoStop}%
\bibitem [{\citenamefont {Visscher}(2002)}]{Visscher2002}%
  \BibitemOpen
  \bibfield  {author} {\bibinfo {author} {\bibfnamefont {L.}~\bibnamefont
  {Visscher}},\ }\href {<Go to ISI>://000175357400002} {\bibfield  {journal}
  {\bibinfo  {journal} {Journal of Computational Chemistry}\ }\textbf {\bibinfo
  {volume} {23}},\ \bibinfo {pages} {759 } (\bibinfo {year}
  {2002})}\BibitemShut {NoStop}%
\bibitem [{\citenamefont {Thyssen}(2001)}]{Thyssen_phd}%
  \BibitemOpen
  \bibfield  {author} {\bibinfo {author} {\bibfnamefont {J.}~\bibnamefont
  {Thyssen}},\ }\emph {\bibinfo {title} {Development and Applications of
  Methods for Correlated Relativistic Calculations of Molecular Properties}},\
  \href@noop {} {Ph.D. thesis},\ \bibinfo  {school} {University of Southern
  Denmark} (\bibinfo {year} {2001})\BibitemShut {NoStop}%
\bibitem [{\citenamefont {Li}\ and\ \citenamefont
  {Paldus}(1997)}]{paldus-externalcorr}%
  \BibitemOpen
  \bibfield  {author} {\bibinfo {author} {\bibfnamefont {X.}~\bibnamefont
  {Li}}\ and\ \bibinfo {author} {\bibfnamefont {J.}~\bibnamefont {Paldus}},\
  }\href@noop {} {\bibfield  {journal} {\bibinfo  {journal} {J. Chem. Phys.}\
  }\textbf {\bibinfo {volume} {107}},\ \bibinfo {pages} {6257} (\bibinfo {year}
  {1997})}\BibitemShut {NoStop}%
\bibitem [{\citenamefont {Lyakh}\ \emph {et~al.}(2011)\citenamefont {Lyakh},
  \citenamefont {Lotrich},\ and\ \citenamefont
  {Bartlett}}]{cyclobut-tailored-2011}%
  \BibitemOpen
  \bibfield  {author} {\bibinfo {author} {\bibfnamefont {D.~I.}\ \bibnamefont
  {Lyakh}}, \bibinfo {author} {\bibfnamefont {V.~F.}\ \bibnamefont {Lotrich}},
  \ and\ \bibinfo {author} {\bibfnamefont {R.~J.}\ \bibnamefont {Bartlett}},\
  }\href@noop {} {\bibfield  {journal} {\bibinfo  {journal} {Chem. Phys.
  Lett.}\ }\textbf {\bibinfo {volume} {501}},\ \bibinfo {pages} {166} (\bibinfo
  {year} {2011})}\BibitemShut {NoStop}%
\bibitem [{\citenamefont {Melnichuk}\ and\ \citenamefont
  {Bartlett}(2012)}]{melnichuk-2012}%
  \BibitemOpen
  \bibfield  {author} {\bibinfo {author} {\bibfnamefont {A.}~\bibnamefont
  {Melnichuk}}\ and\ \bibinfo {author} {\bibfnamefont {R.~J.}\ \bibnamefont
  {Bartlett}},\ }\href@noop {} {\bibfield  {journal} {\bibinfo  {journal} {J.
  Chem. Phys.}\ }\textbf {\bibinfo {volume} {137}},\ \bibinfo {pages} {214103}
  (\bibinfo {year} {2012})}\BibitemShut {NoStop}%
\bibitem [{\citenamefont {Melnichuk}\ and\ \citenamefont
  {Bartlett}(2014)}]{melnichuk-2014}%
  \BibitemOpen
  \bibfield  {author} {\bibinfo {author} {\bibfnamefont {A.}~\bibnamefont
  {Melnichuk}}\ and\ \bibinfo {author} {\bibfnamefont {R.~J.}\ \bibnamefont
  {Bartlett}},\ }\href@noop {} {\bibfield  {journal} {\bibinfo  {journal} {J.
  Chem. Phys.}\ }\textbf {\bibinfo {volume} {140}},\ \bibinfo {pages} {064113}
  (\bibinfo {year} {2014})}\BibitemShut {NoStop}%
\bibitem [{\citenamefont {Piecuch}\ \emph {et~al.}(1993)\citenamefont
  {Piecuch}, \citenamefont {Oliphant},\ and\ \citenamefont
  {Adamowicz}}]{Piecuch1993}%
  \BibitemOpen
  \bibfield  {author} {\bibinfo {author} {\bibfnamefont {P.}~\bibnamefont
  {Piecuch}}, \bibinfo {author} {\bibfnamefont {N.}~\bibnamefont {Oliphant}}, \
  and\ \bibinfo {author} {\bibfnamefont {L.}~\bibnamefont {Adamowicz}},\ }\href
  {\doibase 10.1063/1.466179} {\bibfield  {journal} {\bibinfo  {journal} {The
  Journal of Chemical Physics}\ }\textbf {\bibinfo {volume} {99}},\ \bibinfo
  {pages} {1875} (\bibinfo {year} {1993})}\BibitemShut {NoStop}%
\bibitem [{\citenamefont {Piecuch}\ and\ \citenamefont
  {Adamowicz}(1994)}]{semi3}%
  \BibitemOpen
  \bibfield  {author} {\bibinfo {author} {\bibfnamefont {P.}~\bibnamefont
  {Piecuch}}\ and\ \bibinfo {author} {\bibfnamefont {L.}~\bibnamefont
  {Adamowicz}},\ }\href {https://doi.org/10.1063/1.467143} {\bibfield
  {journal} {\bibinfo  {journal} {The Journal of Chemical Physics}\ }\textbf
  {\bibinfo {volume} {100}},\ \bibinfo {pages} {5792} (\bibinfo {year}
  {1994})}\BibitemShut {NoStop}%
\bibitem [{\citenamefont {Legeza}\ and\ \citenamefont
  {S{\'{o}}lyom}(2003)}]{Legeza2003}%
  \BibitemOpen
  \bibfield  {author} {\bibinfo {author} {\bibfnamefont {{\"O}.}~\bibnamefont
  {Legeza}}\ and\ \bibinfo {author} {\bibfnamefont {J.}~\bibnamefont
  {S{\'{o}}lyom}},\ }\href@noop {} {\bibfield  {journal} {\bibinfo  {journal}
  {Physical Review B}\ }\textbf {\bibinfo {volume} {68}},\ \bibinfo {pages}
  {195116} (\bibinfo {year} {2003})}\BibitemShut {NoStop}%
\bibitem [{\citenamefont {Battaglia}\ \emph {et~al.}(2018)\citenamefont
  {Battaglia}, \citenamefont {Keller},\ and\ \citenamefont
  {Knecht}}]{Battaglia2018}%
  \BibitemOpen
  \bibfield  {author} {\bibinfo {author} {\bibfnamefont {S.}~\bibnamefont
  {Battaglia}}, \bibinfo {author} {\bibfnamefont {S.}~\bibnamefont {Keller}}, \
  and\ \bibinfo {author} {\bibfnamefont {S.}~\bibnamefont {Knecht}},\ }\href
  {\doibase 10.1021/acs.jctc.7b01065} {\bibfield  {journal} {\bibinfo
  {journal} {Journal of Chemical Theory and Computation}\ }\textbf {\bibinfo
  {volume} {14}},\ \bibinfo {pages} {2353} (\bibinfo {year}
  {2018})}\BibitemShut {NoStop}%
\bibitem [{\citenamefont {Szalay}\ \emph {et~al.}(2015)\citenamefont {Szalay},
  \citenamefont {Pfeffer}, \citenamefont {Murg}, \citenamefont {Barcza},
  \citenamefont {Verstraete}, \citenamefont {Schneider},\ and\ \citenamefont
  {Legeza}}]{Szalay2015}%
  \BibitemOpen
  \bibfield  {author} {\bibinfo {author} {\bibfnamefont {S.}~\bibnamefont
  {Szalay}}, \bibinfo {author} {\bibfnamefont {M.}~\bibnamefont {Pfeffer}},
  \bibinfo {author} {\bibfnamefont {V.}~\bibnamefont {Murg}}, \bibinfo {author}
  {\bibfnamefont {G.}~\bibnamefont {Barcza}}, \bibinfo {author} {\bibfnamefont
  {F.}~\bibnamefont {Verstraete}}, \bibinfo {author} {\bibfnamefont
  {R.}~\bibnamefont {Schneider}}, \ and\ \bibinfo {author} {\bibfnamefont
  {{\"O}.}~\bibnamefont {Legeza}},\ }\href {\doibase 10.1002/qua.24898}
  {\bibfield  {journal} {\bibinfo  {journal} {International Journal of Quantum
  Chemistry}\ }\textbf {\bibinfo {volume} {115}},\ \bibinfo {pages} {1342}
  (\bibinfo {year} {2015})}\BibitemShut {NoStop}%
\bibitem [{TUR(2020)}]{TURBOMOLEx}%
  \BibitemOpen
  \href@noop {} {\enquote {\bibinfo {title} {Turbomole v7.52020, a development
  of university of karlsruhe and forschungszentrum karlsruhe gmbh,1989-2007},}\
  } (\bibinfo {year} {2020}),\ \bibinfo {note} {{\tt
  https://www.turbomole.org}}\BibitemShut {NoStop}%
\bibitem [{\citenamefont {Dyall}(2012)}]{dyallbasis}%
  \BibitemOpen
  \bibfield  {author} {\bibinfo {author} {\bibfnamefont {K.~G.}\ \bibnamefont
  {Dyall}},\ }\href@noop {} {\bibfield  {journal} {\bibinfo  {journal} {Theor.
  Chem. Acc.}\ }\textbf {\bibinfo {volume} {131}},\ \bibinfo {pages} {1217}
  (\bibinfo {year} {2012})}\BibitemShut {NoStop}%
\bibitem [{\citenamefont {T.~H.~Dunning}(1989)}]{ccbasis}%
  \BibitemOpen
  \bibfield  {author} {\bibinfo {author} {\bibfnamefont {J.}~\bibnamefont
  {T.~H.~Dunning}},\ }\href@noop {} {\bibfield  {journal} {\bibinfo  {journal}
  {J. Chem. Phys.}\ }\textbf {\bibinfo {volume} {90}},\ \bibinfo {pages} {1007}
  (\bibinfo {year} {1989})}\BibitemShut {NoStop}%
\bibitem [{vib(1979)}]{vibanal}%
  \BibitemOpen
  \href@noop {} {} (\bibinfo {year} {1979}),\ \bibinfo {note} {aNL
  vibration-rotation analysis program for diatomic molecules written by t. h.
  dunning, jr. (winter 1978-9)}\BibitemShut {NoStop}%
\bibitem [{\citenamefont {Balasubramanian}\ \emph {et~al.}(2020)\citenamefont
  {Balasubramanian}, \citenamefont {Chen}, \citenamefont {Coriani},
  \citenamefont {Diedenhofen}, \citenamefont {Frank}, \citenamefont {Franzke},
  \citenamefont {Furche}, \citenamefont {Grotjahn}, \citenamefont {Harding},
  \citenamefont {Hättig}, \citenamefont {Hellweg}, \citenamefont
  {Helmich-Paris}, \citenamefont {Holzer}, \citenamefont {Huniar},
  \citenamefont {Kaupp}, \citenamefont {Marefat~Khah}, \citenamefont
  {Karbalaei~Khani}, \citenamefont {Müller}, \citenamefont {Mack},
  \citenamefont {Nguyen}, \citenamefont {Parker}, \citenamefont {Perlt},
  \citenamefont {Rappoport}, \citenamefont {Reiter}, \citenamefont {Roy},
  \citenamefont {Rückert}, \citenamefont {Schmitz}, \citenamefont {Sierka},
  \citenamefont {Tapavicza}, \citenamefont {Tew}, \citenamefont {van Wüllen},
  \citenamefont {Voora}, \citenamefont {Weigend}, \citenamefont {Wodyński},\
  and\ \citenamefont {Yu}}]{TURBOMOLEx2}%
  \BibitemOpen
  \bibfield  {author} {\bibinfo {author} {\bibfnamefont {S.~G.}\ \bibnamefont
  {Balasubramanian}}, \bibinfo {author} {\bibfnamefont {G.~P.}\ \bibnamefont
  {Chen}}, \bibinfo {author} {\bibfnamefont {S.}~\bibnamefont {Coriani}},
  \bibinfo {author} {\bibfnamefont {M.}~\bibnamefont {Diedenhofen}}, \bibinfo
  {author} {\bibfnamefont {M.~S.}\ \bibnamefont {Frank}}, \bibinfo {author}
  {\bibfnamefont {Y.~J.}\ \bibnamefont {Franzke}}, \bibinfo {author}
  {\bibfnamefont {F.}~\bibnamefont {Furche}}, \bibinfo {author} {\bibfnamefont
  {R.}~\bibnamefont {Grotjahn}}, \bibinfo {author} {\bibfnamefont {M.~E.}\
  \bibnamefont {Harding}}, \bibinfo {author} {\bibfnamefont {C.}~\bibnamefont
  {Hättig}}, \bibinfo {author} {\bibfnamefont {A.}~\bibnamefont {Hellweg}},
  \bibinfo {author} {\bibfnamefont {B.}~\bibnamefont {Helmich-Paris}}, \bibinfo
  {author} {\bibfnamefont {C.}~\bibnamefont {Holzer}}, \bibinfo {author}
  {\bibfnamefont {U.}~\bibnamefont {Huniar}}, \bibinfo {author} {\bibfnamefont
  {M.}~\bibnamefont {Kaupp}}, \bibinfo {author} {\bibfnamefont
  {A.}~\bibnamefont {Marefat~Khah}}, \bibinfo {author} {\bibfnamefont
  {S.}~\bibnamefont {Karbalaei~Khani}}, \bibinfo {author} {\bibfnamefont
  {T.}~\bibnamefont {Müller}}, \bibinfo {author} {\bibfnamefont
  {F.}~\bibnamefont {Mack}}, \bibinfo {author} {\bibfnamefont {B.~D.}\
  \bibnamefont {Nguyen}}, \bibinfo {author} {\bibfnamefont {S.~M.}\
  \bibnamefont {Parker}}, \bibinfo {author} {\bibfnamefont {E.}~\bibnamefont
  {Perlt}}, \bibinfo {author} {\bibfnamefont {D.}~\bibnamefont {Rappoport}},
  \bibinfo {author} {\bibfnamefont {K.}~\bibnamefont {Reiter}}, \bibinfo
  {author} {\bibfnamefont {S.}~\bibnamefont {Roy}}, \bibinfo {author}
  {\bibfnamefont {M.}~\bibnamefont {Rückert}}, \bibinfo {author}
  {\bibfnamefont {G.}~\bibnamefont {Schmitz}}, \bibinfo {author} {\bibfnamefont
  {M.}~\bibnamefont {Sierka}}, \bibinfo {author} {\bibfnamefont
  {E.}~\bibnamefont {Tapavicza}}, \bibinfo {author} {\bibfnamefont {D.~P.}\
  \bibnamefont {Tew}}, \bibinfo {author} {\bibfnamefont {C.}~\bibnamefont {van
  Wüllen}}, \bibinfo {author} {\bibfnamefont {V.~K.}\ \bibnamefont {Voora}},
  \bibinfo {author} {\bibfnamefont {F.}~\bibnamefont {Weigend}}, \bibinfo
  {author} {\bibfnamefont {A.}~\bibnamefont {Wodyński}}, \ and\ \bibinfo
  {author} {\bibfnamefont {J.~M.}\ \bibnamefont {Yu}},\ }\href@noop {}
  {\bibfield  {journal} {\bibinfo  {journal} {The Journal of Chemical Physics}\
  }\textbf {\bibinfo {volume} {152}},\ \bibinfo {pages} {184107} (\bibinfo
  {year} {2020})}\BibitemShut {NoStop}%
\bibitem [{\citenamefont {Furness}\ \emph {et~al.}(2020)\citenamefont
  {Furness}, \citenamefont {Kaplan}, \citenamefont {Ning}, \citenamefont
  {Perdew},\ and\ \citenamefont {Sun}}]{r2scan}%
  \BibitemOpen
  \bibfield  {author} {\bibinfo {author} {\bibfnamefont {J.~W.}\ \bibnamefont
  {Furness}}, \bibinfo {author} {\bibfnamefont {A.~D.}\ \bibnamefont {Kaplan}},
  \bibinfo {author} {\bibfnamefont {J.}~\bibnamefont {Ning}}, \bibinfo {author}
  {\bibfnamefont {J.~P.}\ \bibnamefont {Perdew}}, \ and\ \bibinfo {author}
  {\bibfnamefont {J.}~\bibnamefont {Sun}},\ }\href@noop {} {\bibfield
  {journal} {\bibinfo  {journal} {The Journal of Physical Chemistry Letters}\
  }\textbf {\bibinfo {volume} {11}},\ \bibinfo {pages} {8208} (\bibinfo {year}
  {2020})}\BibitemShut {NoStop}%
\bibitem [{\citenamefont {Holzer}\ \emph {et~al.}(2021)\citenamefont {Holzer},
  \citenamefont {Franzke},\ and\ \citenamefont {Kehry}}]{mpsts-noa2a}%
  \BibitemOpen
  \bibfield  {author} {\bibinfo {author} {\bibfnamefont {C.}~\bibnamefont
  {Holzer}}, \bibinfo {author} {\bibfnamefont {Y.~J.}\ \bibnamefont {Franzke}},
  \ and\ \bibinfo {author} {\bibfnamefont {M.}~\bibnamefont {Kehry}},\
  }\href@noop {} {\bibfield  {journal} {\bibinfo  {journal} {Journal of
  Chemical Theory and Computation}\ }\textbf {\bibinfo {volume} {17}},\
  \bibinfo {pages} {2928} (\bibinfo {year} {2021})}\BibitemShut {NoStop}%
\bibitem [{\citenamefont {Perdew}\ \emph {et~al.}(2008)\citenamefont {Perdew},
  \citenamefont {Staroverov}, \citenamefont {Tao},\ and\ \citenamefont
  {Scuseria}}]{mpsts-noa2b}%
  \BibitemOpen
  \bibfield  {author} {\bibinfo {author} {\bibfnamefont {J.~P.}\ \bibnamefont
  {Perdew}}, \bibinfo {author} {\bibfnamefont {V.~N.}\ \bibnamefont
  {Staroverov}}, \bibinfo {author} {\bibfnamefont {J.}~\bibnamefont {Tao}}, \
  and\ \bibinfo {author} {\bibfnamefont {G.~E.}\ \bibnamefont {Scuseria}},\
  }\href {\doibase 10.1103/PhysRevA.78.052513} {\bibfield  {journal} {\bibinfo
  {journal} {Phys. Rev. A}\ }\textbf {\bibinfo {volume} {78}},\ \bibinfo
  {pages} {052513} (\bibinfo {year} {2008})}\BibitemShut {NoStop}%
\bibitem [{\citenamefont {Andrews}\ \emph {et~al.}(2008)\citenamefont
  {Andrews}, \citenamefont {Wang}, \citenamefont {Lindh}, \citenamefont
  {Roos},\ and\ \citenamefont {Marsden}}]{Andrews2008}%
  \BibitemOpen
  \bibfield  {author} {\bibinfo {author} {\bibfnamefont {L.}~\bibnamefont
  {Andrews}}, \bibinfo {author} {\bibfnamefont {X.}~\bibnamefont {Wang}},
  \bibinfo {author} {\bibfnamefont {R.}~\bibnamefont {Lindh}}, \bibinfo
  {author} {\bibfnamefont {B.}~\bibnamefont {Roos}}, \ and\ \bibinfo {author}
  {\bibfnamefont {C.}~\bibnamefont {Marsden}},\ }\href {\doibase
  https://doi.org/10.1002/anie.200801120} {\bibfield  {journal} {\bibinfo
  {journal} {Angewandte Chemie International Edition}\ }\textbf {\bibinfo
  {volume} {47}},\ \bibinfo {pages} {5366} (\bibinfo {year}
  {2008})}\BibitemShut {NoStop}%
\bibitem [{\citenamefont {VanGundy}(2018)}]{VanGundy_phd}%
  \BibitemOpen
  \bibfield  {author} {\bibinfo {author} {\bibfnamefont {R.~A.}\ \bibnamefont
  {VanGundy}},\ }\emph {\bibinfo {title} {Electronic Structure of
  Metal-Containing Diatomic Ions}},\ \href@noop {} {Ph.D. thesis},\ \bibinfo
  {school} {Faculty of the James T. Laney School of Graduate Studies of Emory
  University} (\bibinfo {year} {2018})\BibitemShut {NoStop}%
\bibitem [{\citenamefont {King}\ and\ \citenamefont
  {Liddle}(2014)}]{KING20142}%
  \BibitemOpen
  \bibfield  {author} {\bibinfo {author} {\bibfnamefont {D.~M.}\ \bibnamefont
  {King}}\ and\ \bibinfo {author} {\bibfnamefont {S.~T.}\ \bibnamefont
  {Liddle}},\ }\href {\doibase https://doi.org/10.1016/j.ccr.2013.06.013}
  {\bibfield  {journal} {\bibinfo  {journal} {Coordination Chemistry Reviews}\
  }\textbf {\bibinfo {volume} {266-267}},\ \bibinfo {pages} {2} (\bibinfo
  {year} {2014})}\BibitemShut {NoStop}%
\bibitem [{\citenamefont {Atkinson}\ \emph {et~al.}(2018)\citenamefont
  {Atkinson}, \citenamefont {Hu},\ and\ \citenamefont
  {Kaltsoyannis}}]{Atkinson2018}%
  \BibitemOpen
  \bibfield  {author} {\bibinfo {author} {\bibfnamefont {B.~E.}\ \bibnamefont
  {Atkinson}}, \bibinfo {author} {\bibfnamefont {H.-S.}\ \bibnamefont {Hu}}, \
  and\ \bibinfo {author} {\bibfnamefont {N.}~\bibnamefont {Kaltsoyannis}},\
  }\href {\doibase 10.1039/C8CC05581E} {\bibfield  {journal} {\bibinfo
  {journal} {Chem. Commun.}\ }\textbf {\bibinfo {volume} {54}},\ \bibinfo
  {pages} {11100} (\bibinfo {year} {2018})}\BibitemShut {NoStop}%
\end{thebibliography}%

\end{document}